\documentclass[twocolumn,times]{aastex62}

\shorttitle{Magnetic Field Extrapolations and Observed Coronal Loops}
\shortauthors{Warren et al.}

\DeclareGraphicsExtensions{.pdf}

\journalinfo{Accepted for Publication in ApJ: April 30, 2018}

\begin{document}


\title{Towards a Quantitative Comparison of Magnetic Field Extrapolations and Observed Coronal
  Loops}

\correspondingauthor{Harry P. Warren}
\email{harry.warren@nrl.navy.mil}

\author[0000-0001-6102-6851]{Harry P. Warren}
\affil{Space Science Division, Naval Research Laboratory, Washington, DC 20375, USA}

\author{Nicholas A. Crump}
\affil{Space Science Division, Naval Research Laboratory, Washington, DC 20375, USA}

\author[0000-0001-5503-0491]{Ignacio Ugarte-Urra}
\affil{Space Science Division, Naval Research Laboratory, Washington, DC 20375, USA}

\author[0000-0003-4043-616X]{Xudong Sun}
\affil{Institute for Astronomy, University of Hawaii at Manoa, Pukalani, HI 96768, USA}

\author[0000-0003-0260-2673]{Markus J. Aschwanden} \affil{Lockheed Martin, Solar and Astrophysics
  Laboratory, Org. A021S, Bldg. 252, 3251 Hanover Street, Palo Alto, CA 94304, USA}

\author{Thomas Wiegelmann} \affil{Max-Planck-Institut f{\"u}r Sonnensystemforschung,
  Justus-von-Liebig-Weg 3, 37077 G{\"o}ttingen, Germany}


\begin{abstract}
It is widely believed that loops observed in the solar atmosphere trace out magnetic field
lines. However, the degree to which magnetic field extrapolations yield field lines that actually
do follow loops has yet to be studied systematically. In this paper we apply three different
extrapolation techniques --- a simple potential model, a NLFF model based on photospheric vector
data, and a NLFF model based on forward fitting magnetic sources with vertical currents --- to 15
active regions that span a wide range of magnetic conditions. We use a distance metric to assess
how well each of these models is able to match field lines to the 12,202 loops traced in coronal
images. These distances are typically 1--2\arcsec. We also compute the misalignment angle between
each traced loop and the local magnetic field vector, and find values of 5--12$^\circ$. We find
that the NLFF models generally outperform the potential extrapolation on these metrics, although
the differences between the different extrapolations are relatively small. The methodology that we
employ for this study suggests a number of ways that both the extrapolations and loop
identification can be improved.

\end{abstract}

\keywords{Sun: corona}


\section{Introduction} \label{sec:intro}

It is universally accepted that magnetic fields play a critical role in a wide range of solar
{phenomena}, such as the heating of the solar upper atmosphere, the origin of the solar wind, and
the initiation of coronal mass ejections. Unfortunately, at present, accurate magnetic field
measurements over a wide field of view are routinely available only in the solar photosphere, where
the magnetic field is still dominated by the plasma pressure and the field is not force-free
\citep[e.g.,][]{metcalf1995}. This greatly complicates the use of photospheric measurements as a
boundary condition for {methods that use the force-free assumption to extrapolate the magnetic
  field into the solar chromosphere and corona.}

One approach to addressing the mismatch between the photospheric measurements and the force-free
condition is to preprocess the vector magnetic field observations so that they approximate what
would be measured in the chromosphere, where the field does become force-free. {The non-linear
  force-free (NLFF) code introduced by \citet{wiegelmann2006} represents perhaps the most widely
  known example of this approach. Other codes implementing this idea include those presented by
  \citet{valori2012}, and \citet{jiang2013}. These codes build on the earlier NLFF models of
  \citet{wiegelmann2004}, \citet{amari2006}, and \citet{wheatland2007}.}

Recently, an alternative approach to modeling non-potential fields in the corona has been developed
that does not rely on vector magnetic field measurements. Instead, the line-of-sight photospheric
magnetic field is modeled as a superposition of magnetic sources. The currents associated with each
of these sources is varied in order to optimize the agreement between loops traced in coronal
images and the topology of the field (see, \citealt{aschwanden2013a, aschwanden2013b,
  aschwanden2016}; also see \citealt{malanushenko2012} for a variation on this approach).

Models of the magnetic field play a critical role in our ability to study coronal heating. For
example, a number of studies have used magnetic field extrapolations to determine the relationship
between heating rates and the properties of the field by comparing full active region hydrodynamic
simulations to observations \citep[e.g.,][]{schrijver2004, warren2007, lundquist2008,
  winebarger2008, bradshaw2016,ugarte-urra2017}.  These studies have often found that a volumetric
heating rate that scales approximately as $\bar{B}/L$, where $\bar{B}$ is the mean field strength
and $L$ is the loop length, provides a good match between the simulation and the global properties
of the observed active region.

Interestingly, the alternative approach, where measurements of intensity variations on individual
loops or plasma parameters on individual loops are related to the properties of the associated
field lines appears to have received relatively little attention (see \citealt{xie2017} for one
such example). This is surprising given that the trend in solar instrumentation is towards higher
spatial and temporal resolution. The High-Resolution Coronal Imager (Hi-C; \citealt{kobayashi2014})
and the Interface Region Imaging Spectrograph (IRIS; \citealt{depontieu2014}), for example, achieve
a spatial resolution of better than 360\,km and cadences below 10\,s.

One impediment to studying the relationship between the properties of individual loops and the
properties of magnetic field lines is the difficultly of matching the two together. Since loops are
projected onto a two dimensional plane, their three dimensional geometry is ambiguous, except in
the rare case of stereoscopic observations. Perhaps more fundamentally, it is not clear how closely
current extrapolation techniques reproduce the topological properties of the corona, and how
force-free the corona is at each loop location. Coronal images certainly show many clear examples
of loops, or at least partial segments of loops, but systematic comparisons between the different
extrapolation techniques and these loops have yet to be carried out. Systematic studies of
different NLFF extrapolation methods have generally focused on the more global properties of the
field, such as the free energy or the helicity (see, for example, \citealt{derosa2009} and
\citealt{derosa2015}). {Some previous studies have compared extrapolated field lines to loops
  reconstructed from stereoscopic observations, \citet{derosa2009} and \citet{chifu2017}, but these
  comparisons have been limited to only a few loops.}

In this paper we perform systematic comparisons of several extrapolation techniques with loops
traced in coronal images. We consider the Vertical-Current Approximation (VCA) NLFF method
described in \citet{aschwanden2016} and the NLFF extrapolation method based on vector observations
described in \citet{wiegelmann2012}. For reference, we also consider a simple potential field
extrapolation. Observations of the photospheric field are taken from the Helioseismic and Magnetic
Imager (HMI, \citealt{scherrer2012}) on the \textit{Solar Dynamics Observatory} (SDO). We apply all
three methods to the 15 active regions analyzed in \citet{warren2012}, which represent a broad
spectrum of active regions sizes and total magnetic fluxes. For each of these regions we trace
loops in coronal images taken with the Atmospheric Imaging Assembly (AIA/SDO, \citealt{lemen2012}),
using an established technique \citep{aschwanden2010, aschwanden2013c}. To evaluate the
extrapolations we consider two metrics: the mean distance between the projected field line and
the traced loops and the misalignment angle between the local field vector and the traced loops.

We find that the NLFF models generally outperform the potential extrapolation, although the
differences are relatively small. The vector NLFF code produces smaller mean distances between the
best-fit field line and the traced loops than the potential extrapolation. The VCA code produces
smaller misalignment angles between the traced loops and the local field vector than the potential
extrapolation.

\begin{figure*}[t!]
\centerline{
  \includegraphics[width=0.975\textwidth]{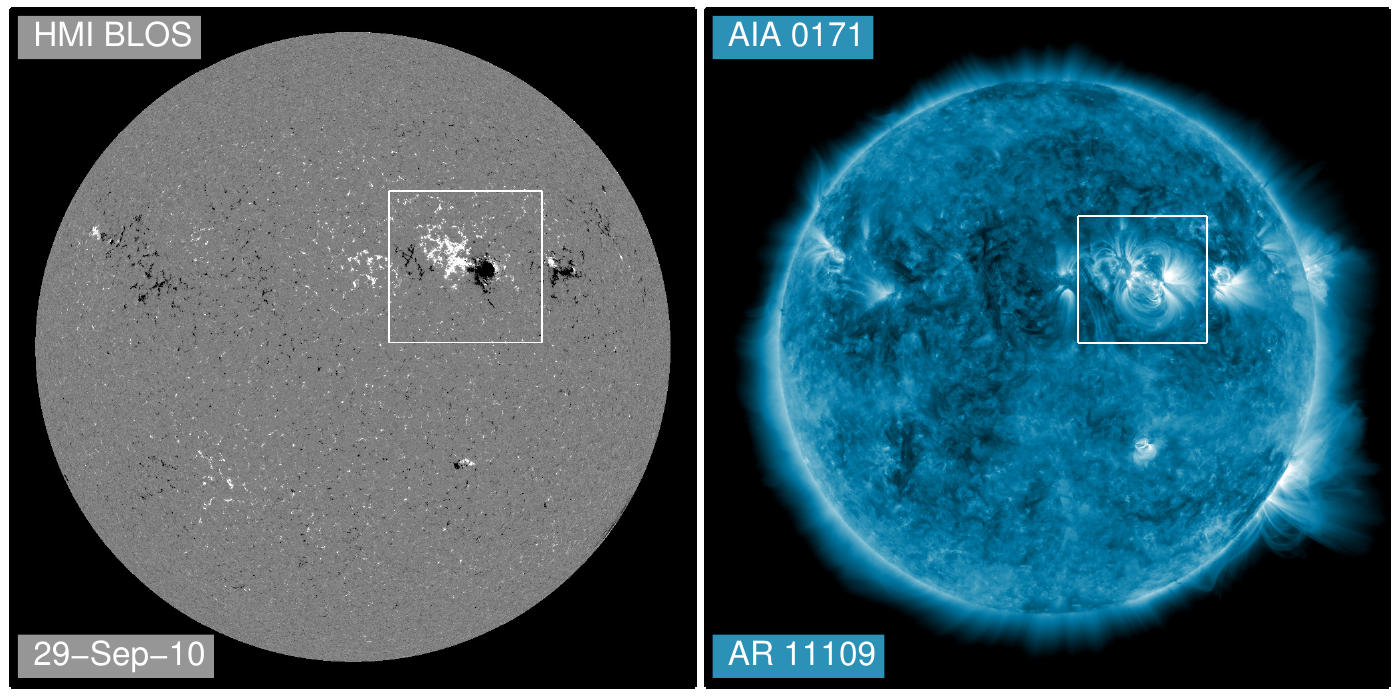}}
\caption{Left: SDO/HMI line-of-sight magnetogram. Right: SDO/AIA 171 \AA\ image. Both images are
  from 29 September 2010 at 23:21:34. The box in each image shows the patch around active region
  11109 selected for the magnetic field extrapolations.}
\label{fig:fov}
\end{figure*}

The objective of this paper is to assess how well these existing extrapolation methods reproduce
the observed topology of the corona. A future paper will focus on comparing the properties of
transient heating events to the properties of the underlying field lines.

This paper is structured in the following way. In Section~\ref{sec:mfe} we provide a brief overview
of the different magnetic field extrapolation methods and the field line calculations. In
Section~\ref{sec:loops} we describe the loop tracing and the methods for matching traced loops to
field lines. The results from applying this methodology to over 12,000 loops sampled from the 15
active regions is presented in Section~\ref{sec:res}. A summary and discussion, including a
discussion of possible improvements to both the extrapolation methods and loop identification, are
presented in Section~\ref{sec:summary}.
\section{Magnetic Field Extrapolations} \label{sec:mfe}

\begin{figure}[t!]
  \vspace{2mm}
  \centerline{%
    \includegraphics[width=0.95\linewidth]{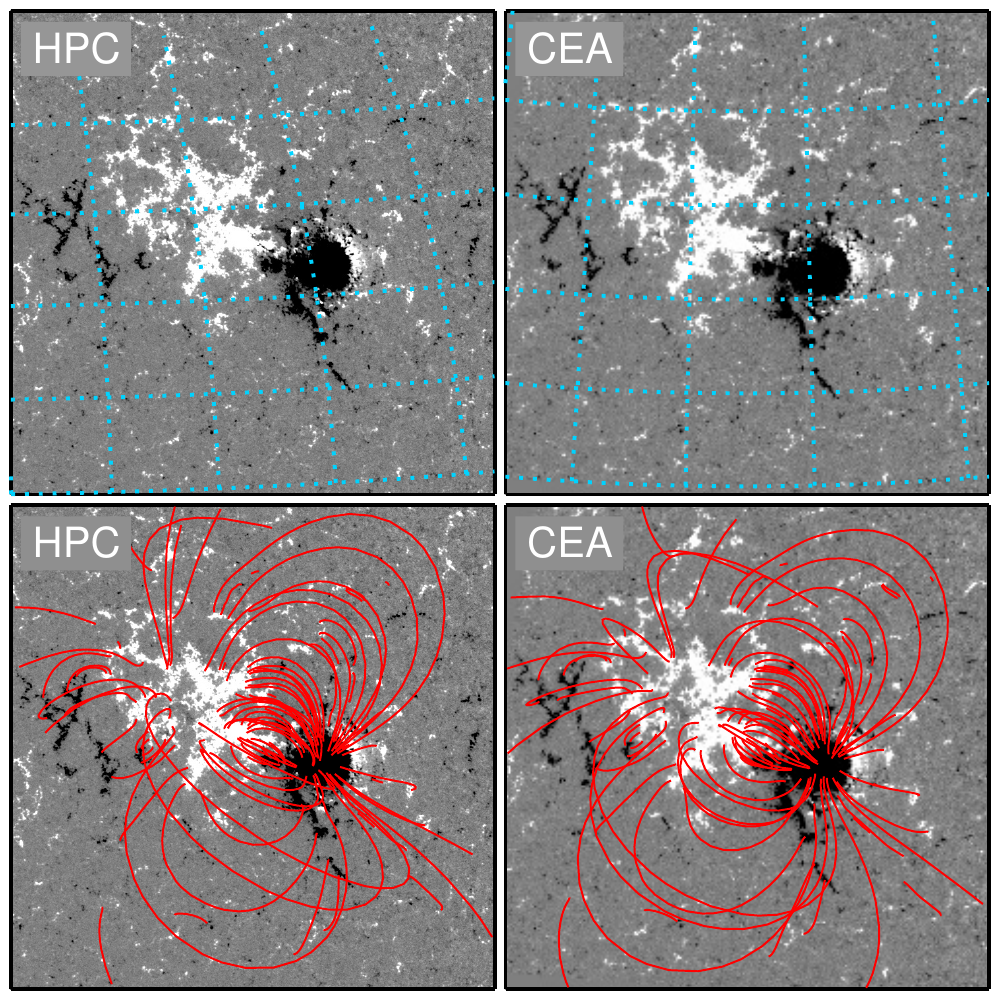}}
  \caption{HMI line-of-sight magnetogram patch for active region 11109 on 29 September 2010.  Top
    left: the native magnetogram in helioprojective Cartesian coordinates (HPC).  Top right: the
    magnetogram transformed into the cylindrical equal-area projection (CEA). Lines of heliographic
    latitude and longitude are shown in each of the top images. Bottom right: field
    lines from the potential field extrapolation computed in the CEA coordinates and overplotted on
    the CEA image. Bottom left: the same field lines projected back onto the native HPC
    magnetogram.}
\label{fig:hpc2cea}
\end{figure}

\subsection{Potential Field Extrapolation}

A potential extrapolation is a solution to the equations
\begin{equation}
    \nabla\times\mathbf{B} = 0,\quad \nabla\cdot\mathbf{B} = 0
\end{equation}

using the corrected line of sight component of the observed magnetic field as the lower boundary
condition. Introducing the scalar potential $\mathbf{B} = \nabla\varphi$ reduces this to solving
Laplace's equation, $\nabla^2\varphi=0$. For this work we use a modified version of a solver based
on Fourier transforms \citep[e.g.,][]{alissandrakis1981} that we have used in previous studies
\citep[e.g.,][]{warren2006,ugarte-urra2007,ugarte-urra2017}. One of the modifications for this work
is to project the observed field from the helioprojective Cartesian (HPC) coordinate system in
which the data are taken to a cylindrical equal area (CEA) coordinate system\footnote{See
  \citet{thompson2006} for a detailed discussion of various coordinate systems and the
  transformations between them.} \citet{sun2013} provides additional details on the transformations
to and from a CEA projection. This corrects for foreshortening in the observed magnetograms. See
Figures~\ref{fig:fov} and \ref{fig:hpc2cea} for an example of such a projection applied to an
observation.  Previously we had considered regions close to disk center where foreshorting effects
are smaller. We also pad the perimeter so that field lines close locally rather than connecting to
sources in ``adjacent'' regions of the periodic domain. This calculation yields the magnetic field
components on a Cartesian grid in the CEA coordinate system. We typically use a grid spacing of
about $0.1^\circ$ per pixel or about $1.8\arcsec$ per pixel and extrapolate up to a height equal to
a box side, typically several hundred arc-seconds. A typical calculation takes about 60\,s on a
standard workstation.

\subsection{Wiegelmann Non-Linear Force Free}

A non-linear force-free magnetic field is a solution to the equations
\begin{equation}
  \nabla\times\mathbf{B} = \frac{4\pi}{c}\mathbf{J} = \alpha\mathbf{B},\quad \nabla\cdot\mathbf{B}
  = 0,
  \label{eq:nlff}
\end{equation}
so that $\mathbf{J}\times\mathbf{B} = 0$.  The twist parameter $\alpha$ is constant along each
field line, but varies from field line to field line. As mentioned previously, the lower boundary
condition is derived by preprocessing the observed photospheric vector field measurements to make
them more consistent with the force-free assumption \citep{wiegelmann2006}. The preprocessing
attempts to find a modified version of the field that is free of forces and torques, is relatively
smooth, but is also close to what is observed. The field components at each point are determined by
starting with the observations and using gradient descent to find the optimal balance between the
four conditions, given a set of relative weights specified by the user.

Once the boundary conditions are determined, the field components are determined by minimizing a
functional that is the sum of the squares of the terms in Equation~\ref{eq:nlff} and an error term
(see \citealt{wiegelmann2012} Equation 4). As with the potential extrapolation, the NLFF uses the
observed photospheric field projected into a CEA coordinate system \citep[see][for details on the
projection of the vector components]{sun2013}. We use a resolution of about $0.06^\circ$ per pixel
or about $1.0\arcsec$ per pixel and extrapolate up to a height of about $160\arcsec$. A typical
calculation takes about 6 hours on a standard workstation.

\subsection{Aschwanden Vertical Current Forward Fit}

The VCA-NLFF method of \citet{aschwanden2016} uses the radial component of the observed magnetic
field as the lower boundary condition and assumes that the field can be represented as a linear
superposition of sub-photospheric sources of the form
\begin{eqnarray}
  B_r & = & B_0\left(\frac{d^2}{r^2}\right)\frac{1}{1+b^2r^2\sin^2\theta}, \\
  B_\phi & = & B_0\left(\frac{d^2}{r^2}\right)\frac{br\sin\theta}{1+b^2r^2\sin^2\theta}, \\
  B_\theta & = & 0,
\end{eqnarray}
where $(r,\theta,\phi)$ are spherical coordinates centered at the magnetic source and $d$ is the
depth below the photospheric surface. The parameter $b$ is related to the twist of the field by
\begin{equation}
\alpha = \frac{2b\cos\theta}{(1 + b^2r^2\sin^2\theta)}.
\end{equation}
Note that this representation for the magnetic field is force-free and divergence-free to second
order (in the parameter $\alpha$ or $b$, {for small values of $\alpha$ or $b$}).  It is
analytically shown that the VCA-NLFFF approximation is exactly divergence-free and force-free in
the vertical loop segments near the loop axis above each buried magnetic charge (see
\citealt{aschwanden2013a} Section 3.3).

The twist parameters for the magnetic sources are iteratively adjusted to minimize the misalignment
angle {defined in Section~\ref{sec:angle}}, that is, to provide the best match between the local
magnetic field vector and the loops traced in coronal images. Note that to determine the position
of the observed loop in three-dimensional space, which is necessary to compute the magnetic field
vector, the loop is fit as a circular loop segment, {which drives the optimization of the alpha
  values in the final computed field lines.} It is not compared with a computed field
line. {Also, the loops that we use to tune the parameters in the VCA-NLFF method and the
  loops that we use to benchmark it are derived from the same procedure (the OCCULT code). We will
  return to these issues in the next sections.}

The resulting extrapolation yields the radial and azimuthal components of the field. These vector
components are transformed to Cartesian coordinates, which is consistent with the outputs of the
other codes. This method rebins the input magnetogram to a resolution of about $1.5\arcsec$ per
pixel and extrapolates up to a height of several hundred arc-seconds. A typical calculation takes
about 10 minutes to converge on a standard workstation. The VCA-NLFF code is written in IDL and
distributed through SolarSoftWare (SSW, \citealt{freeland1998}).

\subsection{Computing Field Lines}

Each of the extrapolation techniques yields the components of the magnetic field in a Cartesian
coordinate system. Thus the field lines are determined by
\begin{equation}
\frac{ds}{B} = \frac{dx}{B_x} = \frac{dy}{B_y} = \frac{dz}{B_z},
\end{equation}
which we integrate using a fourth-order Runge-Kutta method with adaptive step size
\citep{press1992}. To accelerate the calculation of field lines we have written this code in C.

For the potential and vector NLFF methods we need to project the field lines computed in the CEA
coordinate system back to the helioprojective Cartesian coordinate system of the original
image. This transformation is a multi-step process. We first transform $(x_{cea}, y_{cea})$ to
heliographic latitude and longitude $(\theta, \varphi)$ and then extend these coordinates to
Stonyhurst heliographic coordinates $(r, \theta, \varphi)$ assuming $r = R_\odot + z_{cea}$. These
coordinates are then transformed to helioprojective-Cartesian, which accounts for the apparent
latitude (B-angle) and longitude of the observation at Earth (see Equation 11 of
\citealt{thompson2006}). An example of this mapping of field lines from CEA to HPC is shown in
Figure~\ref{fig:hpc2cea}. The transformation from the three-dimensional Cartesian box of the
extrapolation to the spherical geometry of the sun is not unique and introduces unavoidable
distortions that increase with height away from the surface. The VCA-NLFF extrapolation is computed
in Stonyhurst coordinates, so only the final transformation to helioprojective-Cartesian
coordinates is needed to map field lines back to the image plane. Finally, we note that the image
time need not be close to the time of the magnetogram used for the extrapolation. It is a simple
matter to rotate the coordinates of the field lines in heliographic coordinates.
\begin{figure*}[t!]
  \centerline{\includegraphics[width=0.975\textwidth]{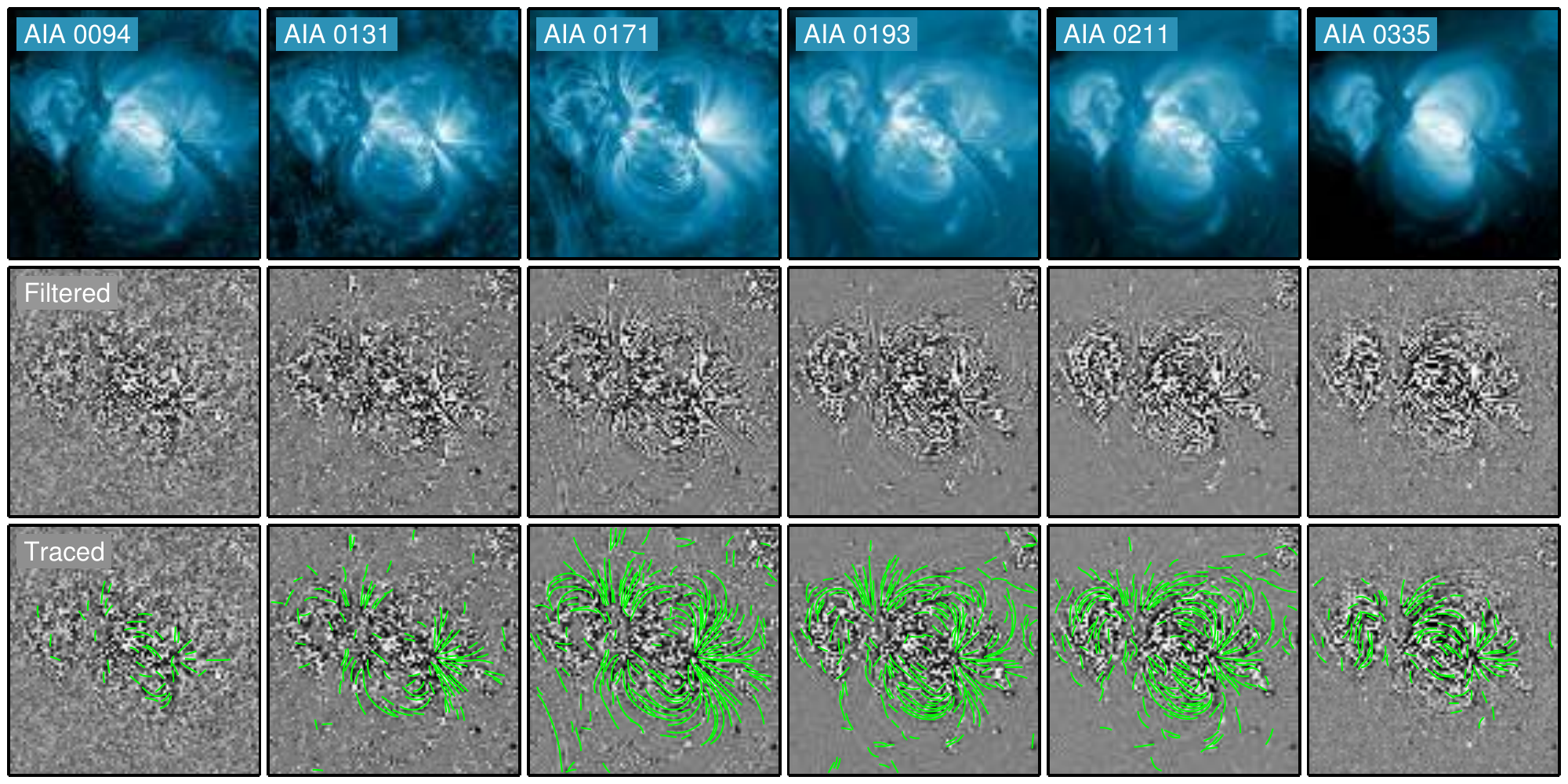}}
  \caption{An example of observed coronal loops identified using automated loop tracing on AIA
    images from AR 11109.  The top row shows a 8-image average for each AIA channel used. Images in
    the middle row are high-pass/low-pass filtered to identify high signal-to-noise features.
    Automatically identified loop segments from each image are displayed in green in the bottom
    row.}
  \label{fig:tracing}
\end{figure*}

\section{Comparison to Coronal Loops} \label{sec:loops}

\subsection{Automated Loop Identification} \label{sec:auto}

To make systematic comparisons between many loops traced in coronal images and the field lines
computed from the magnetic field extrapolations, we must use an automated loop tracing
algorithm. Methods for comparing field lines to observed coronal structures without explicitly
tracing the loops have been developed \citep[e.g.,][]{carcedo2003,conlon2010}, but these approaches
require user inputs for each case. For our work we use the Oriented Coronal CUrved Loop Tracing
(OCCULT) code described in \citet{aschwanden2010} and \citet{aschwanden2013c}. The first step in
this algorithm is to compute the difference between lowpass and highpass filtered versions of an
image. This eliminates both the large-scale background and noise. The second step is to trace along
the intensity ridge emanating from the brightest point in the image. After a loop is identified,
the pixels in the image associated with it are set to zero and the process is repeated.

The OCCULT code is used in the VCA-NLFF algorithm to identify loop segments. We, however, also run
it as a separate module on data that we have processed independently. The VCA code automatically
downloads a single, full-disk AIA image for each wavelength of interest. We found that by
downloading a {time sequence} of AIA cutouts for the region of interest and averaging them together
we are able to identify a larger number of loops.  Loops traced in the averaged images also tend to
be longer and appear to be more complete. An example of loops traced on a set of 6 AIA EUV images
from AR 11109 is shown in Figure~\ref{fig:tracing}.

\subsection{Mapping Field Lines to Traced Loops}

\begin{figure*}
  \centerline{%
    \includegraphics[width=0.32\textwidth]{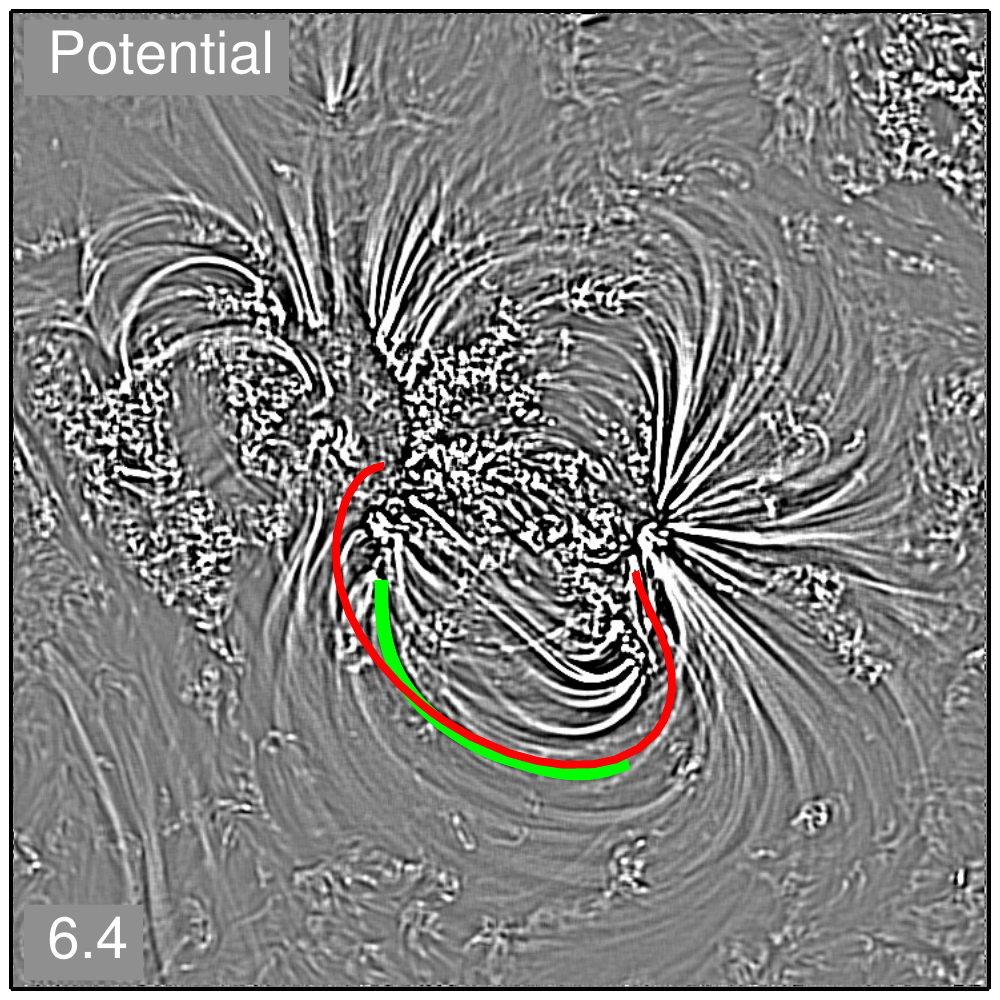}
    \includegraphics[width=0.32\textwidth]{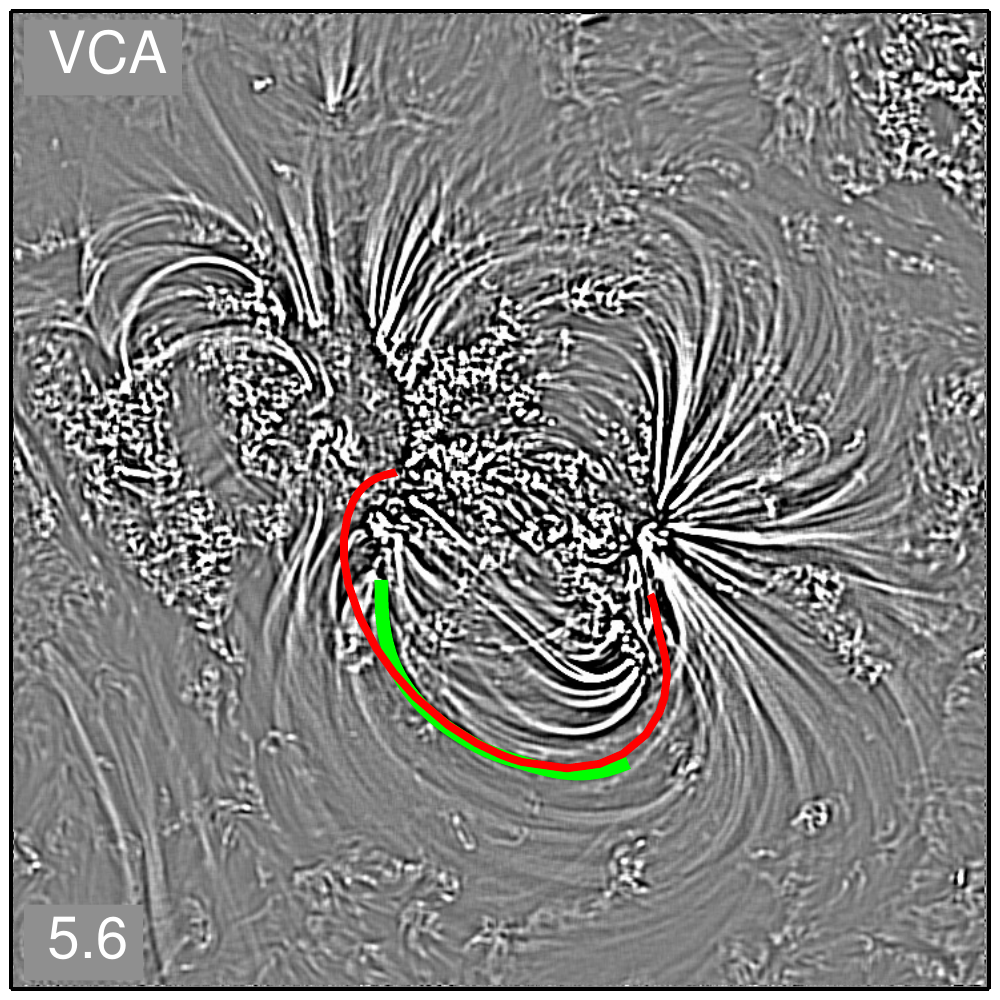}
    \includegraphics[width=0.32\textwidth]{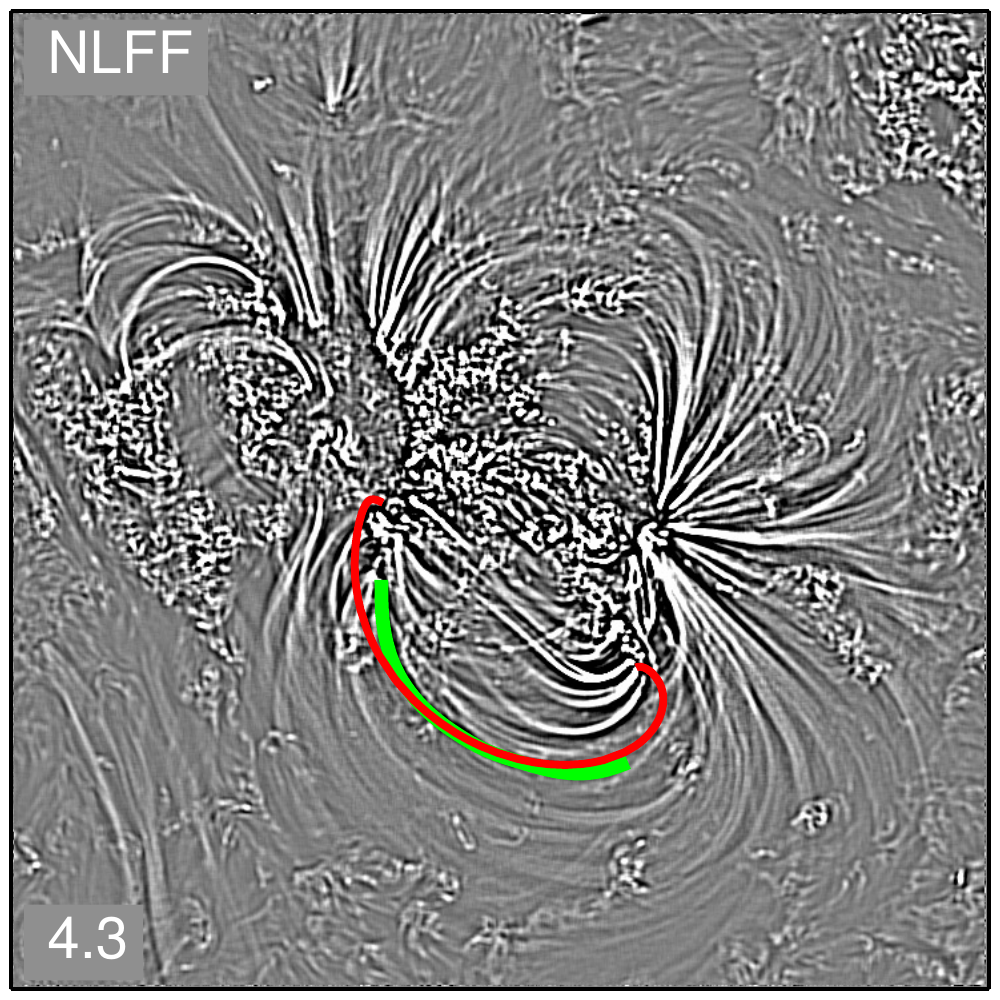}
  }
  \centerline{%
    \includegraphics[trim=24 0 26 90bp,clip,width=0.50\textwidth]{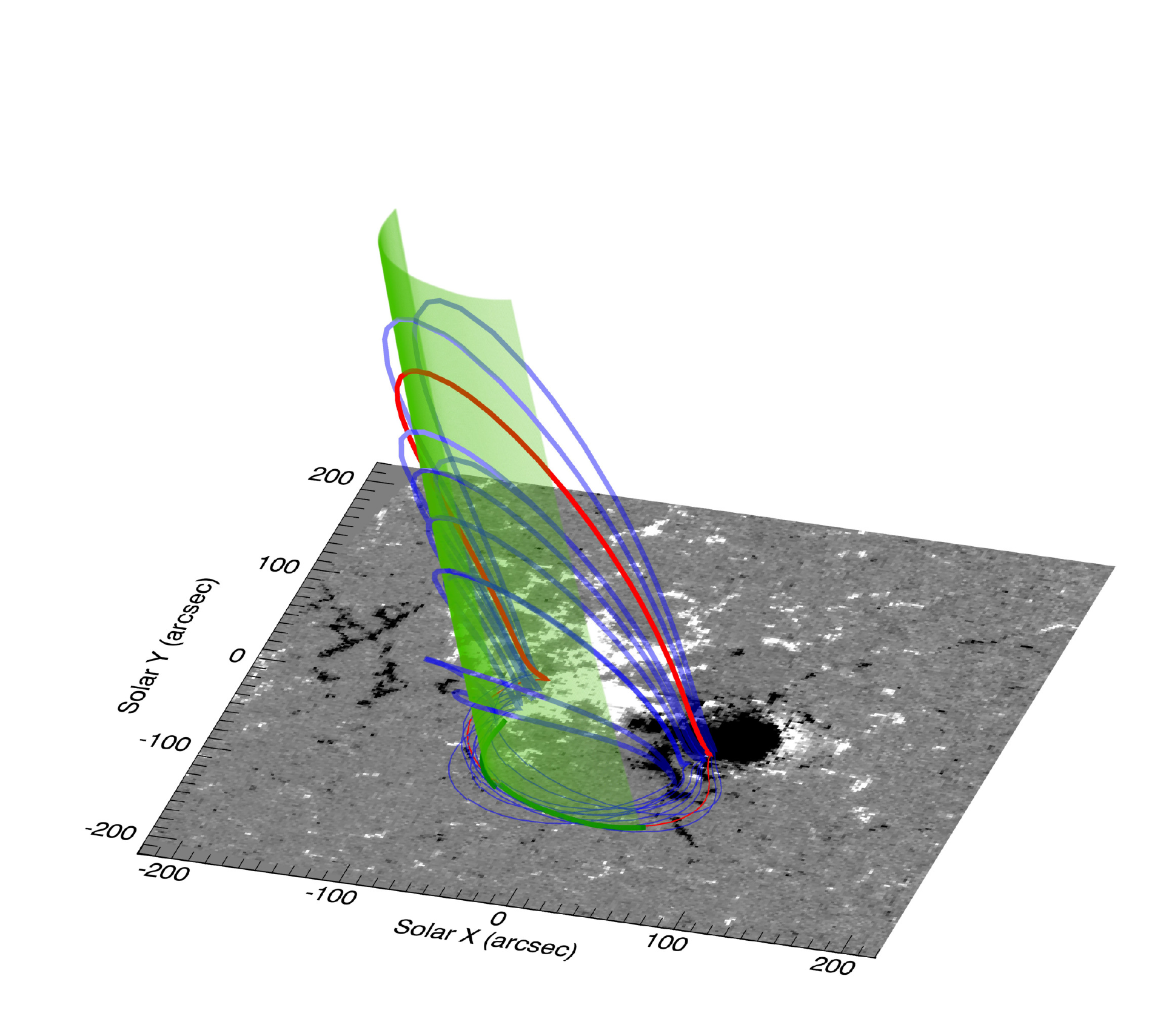}
    \includegraphics[trim=50 0 0 90bp,clip,width=0.50\textwidth]{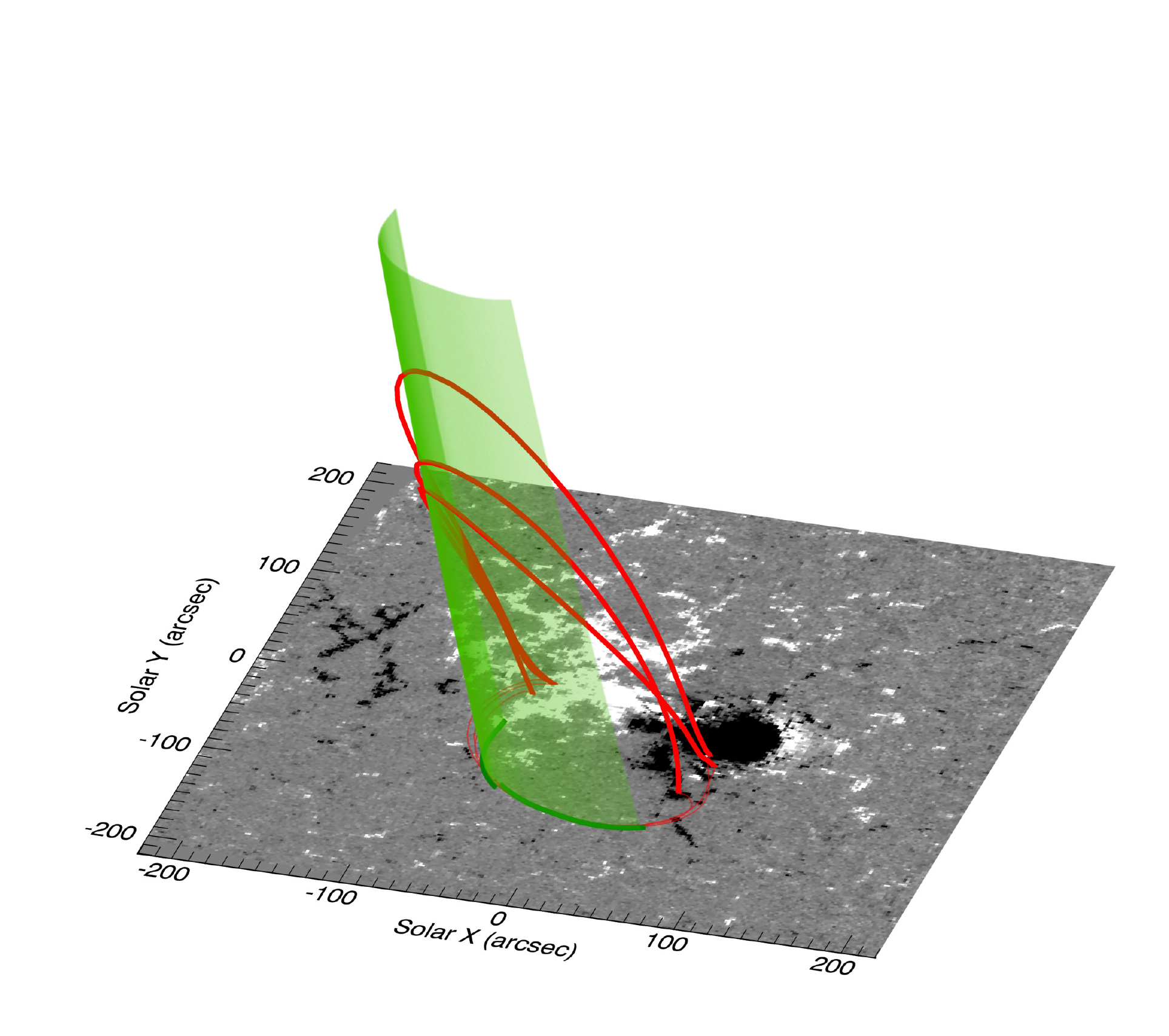}
  }
  \caption{Best-fit field lines for an observed loop in active region 4.  Top panels: an observed
    coronal loop (green) in AIA 171 \AA\ along with the best-fit field line (red) from each
    extrapolation. The mean minimum distance values are given in the lower left corner of each
    image. Bottom left panel: To find the best-fit field line we project the traced loop back to
    the coordinate system of the extrapolation assuming a range of possible heights. Here we show
    the potential case. This projection forms a surface or ``curtain'' (shown in light green),
    which is used to generate seed points for computing candidate field lines (shown in blue). The
    candidate field lines are projected onto the image plane for comparison with the traced
    loop. The field line that produces the smallest mean distance is considered to be the best-fit
    field line (shown in red). This process is repeated for all three extrapolations. Bottom right
    panel: The best-fit field lines from the three extrapolations plotted in the CEA coordinate
    system.}
  \label{fig:curtain1}
\end{figure*}

\begin{figure*}
  \centerline{%
    \includegraphics[width=0.32\textwidth]{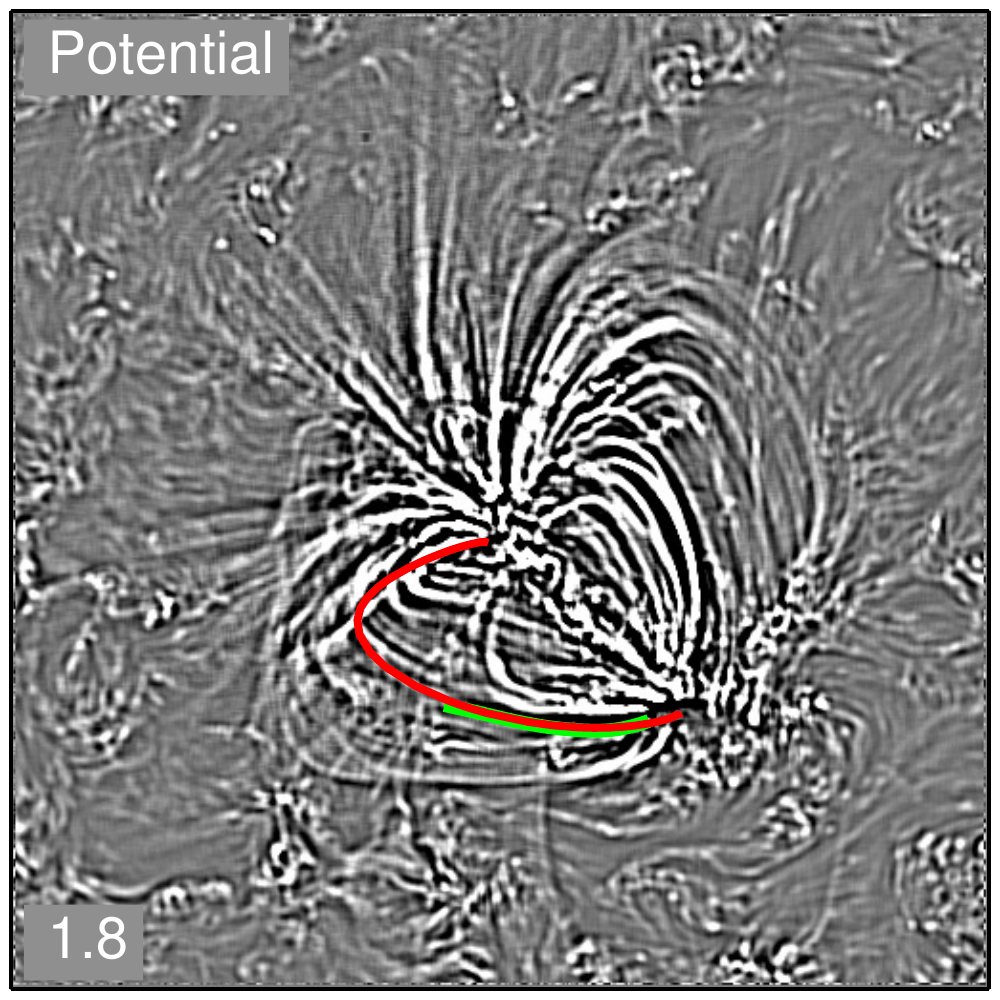}
    \includegraphics[width=0.32\textwidth]{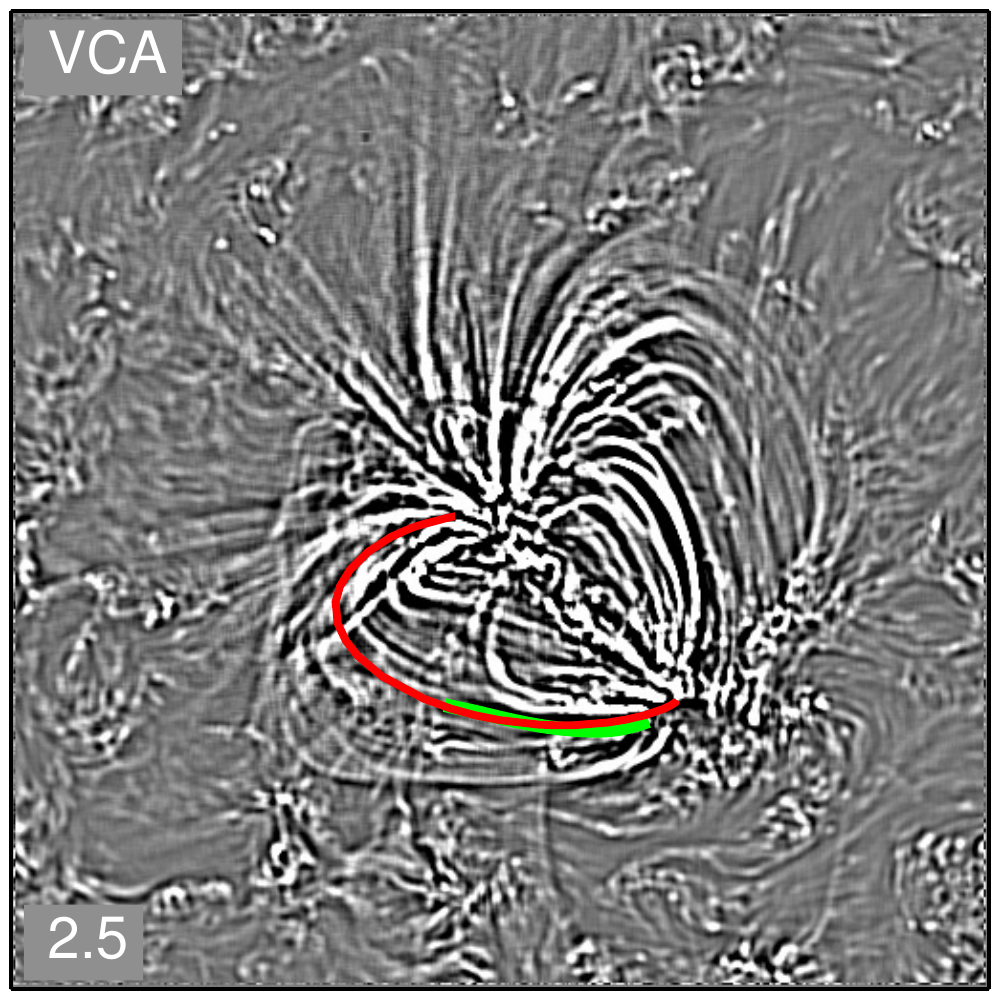}
    \includegraphics[width=0.32\textwidth]{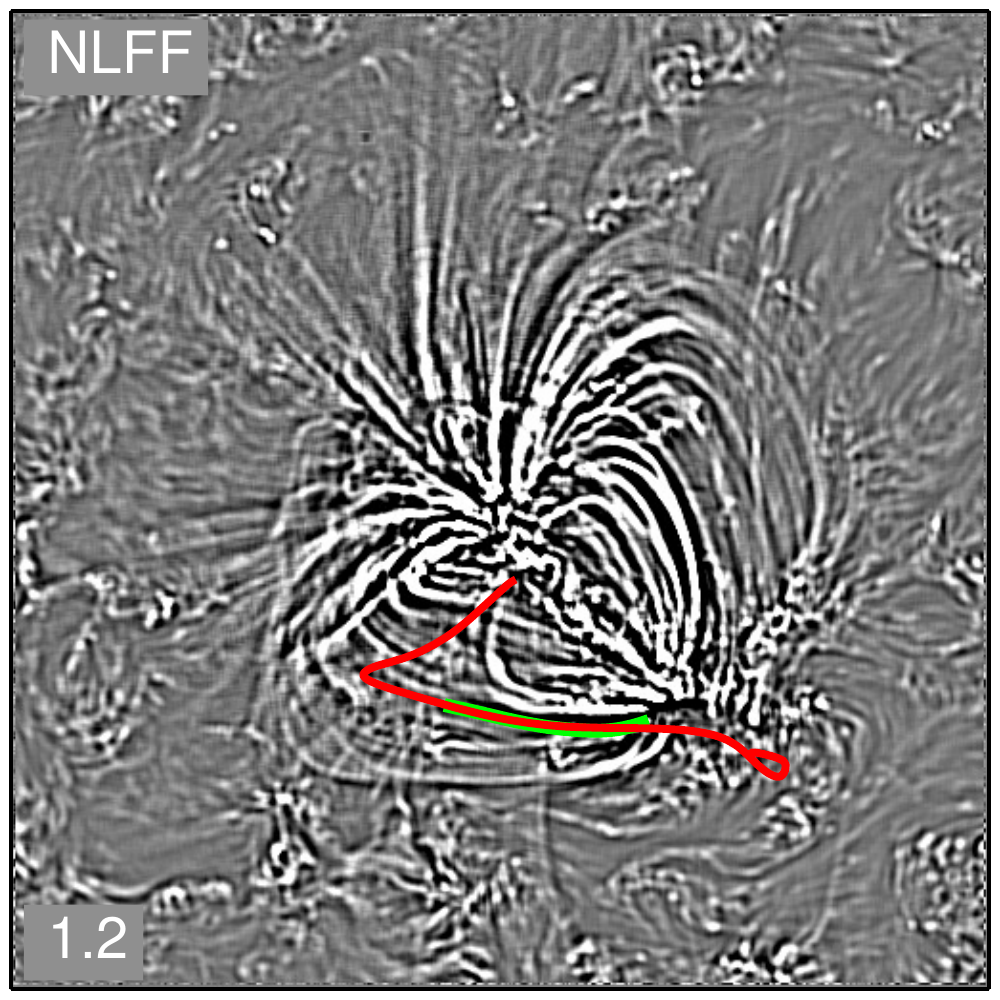}
  }
  \centerline{%
    \includegraphics[trim=24 0 26 160bp,clip,width=0.50\textwidth]{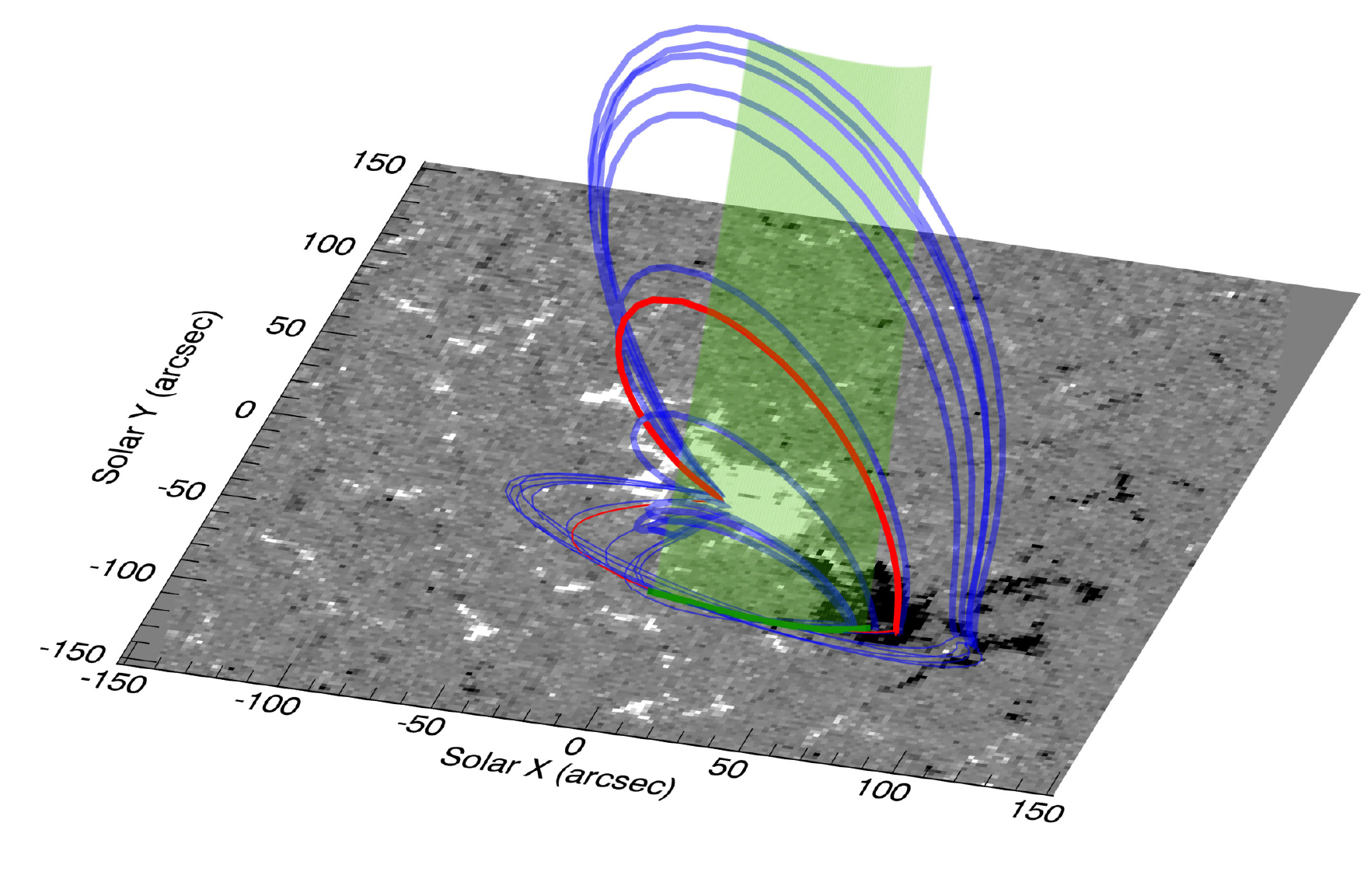}
    \includegraphics[trim=50 0 0 160bp,clip,width=0.50\textwidth]{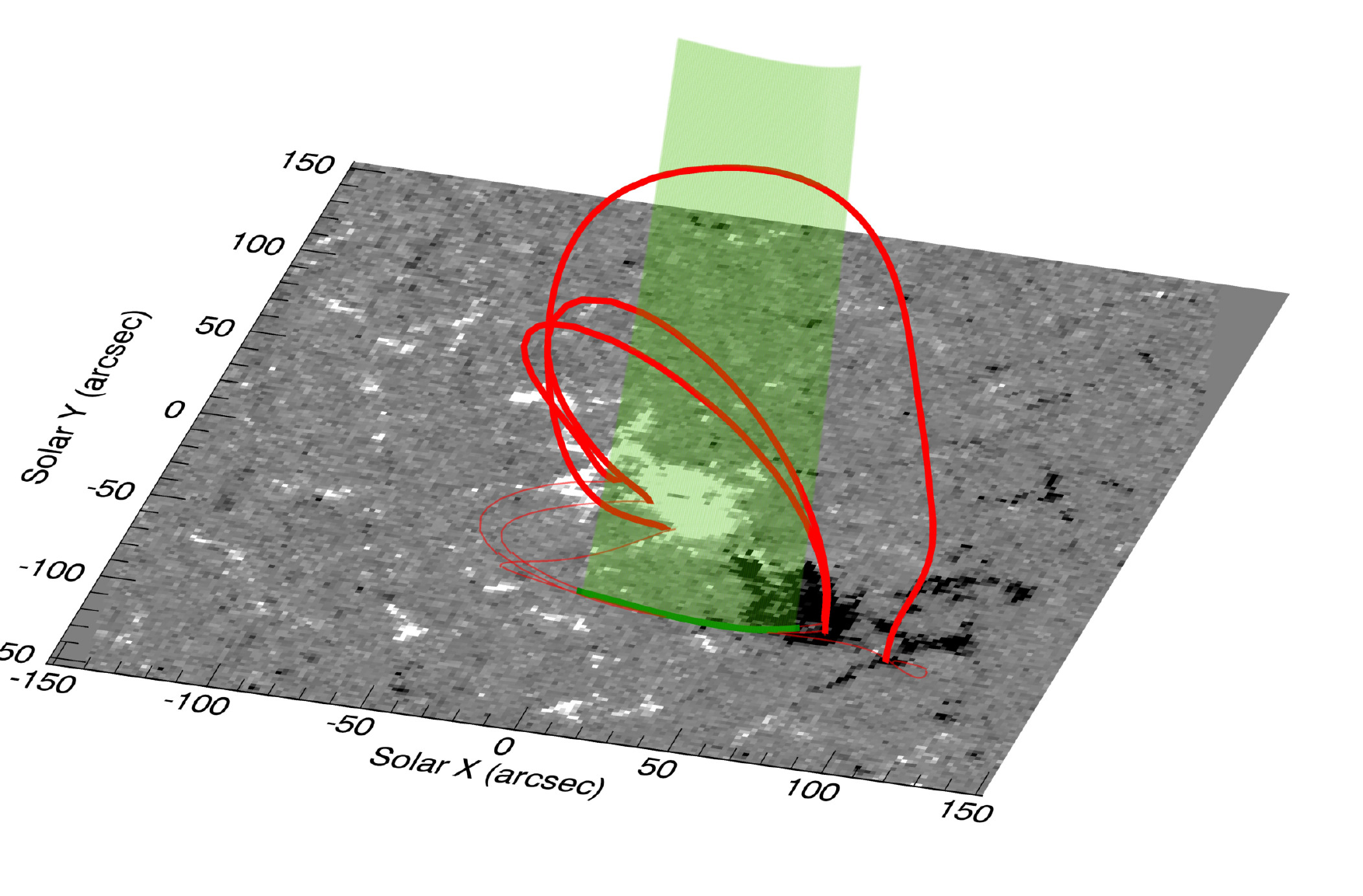}
  }
  \caption{Same as Figure~\ref{fig:curtain1} for a loop in active region 1. This example
    illustrates the differences in the magnetic field topology produced by the three models.}
  \label{fig:curtain2}
\end{figure*}

\begin{figure}[h!]
  \centerline{\includegraphics[width=0.45\textwidth]{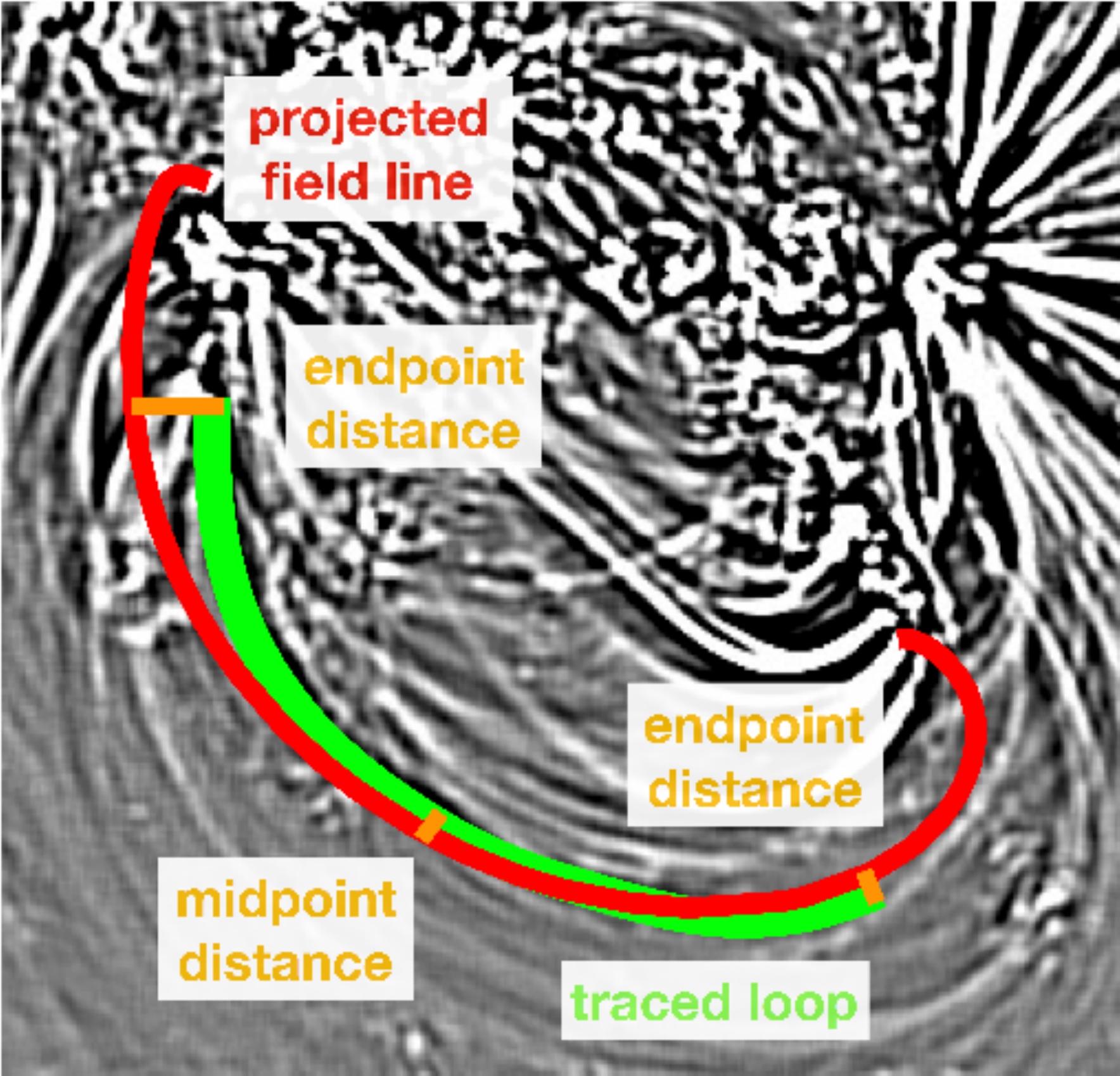}}
  \caption{An example of a mean minimum distance calculation with $N=3$. The shortest distance
    between the endpoints and the midpoint of the traced loop (shown in green) and any point on the
    projected field line (shown in red) are averaged. This metric is used to evaluate how closely
    projected field lines match traced loops. For this work we scale $N$ with loop length and
    between 4 and 50 points are considered.}
  \label{fig:mmd}
\end{figure}

The final element of this program is a method for matching each traced loop in the AIA images to a
field line computed from the extrapolated magnetic field. As mentioned previously, the principal
problem is that the traced loops are projected onto the image plane and the three dimensional
geometry of the loop is ambiguous. To overcome this we map the 2D coordinates of the traced loop
back to the 3D geometry of the extrapolation assuming a range of possible heights. This maps the
one dimensional traced loop to a two dimensional, curtain-like surface. Points on this surface are
then used as initial conditions for calculating field lines. These field lines are then projected
back onto the image plane where they can be compared with the traced loop. Examples of this
calculation are shown in Figures~\ref{fig:curtain1} and \ref{fig:curtain2}.

To determine how well matched a given field line is to a traced loop, we compute the average of the
minimum Euclidean distance between several points on the traced loop and the field line. We refer
to this average as the ``mean minimum distance.'' Mathematically this is
\begin{equation}
  \mathrm{Mean\ Minimum\ Distance} = \frac{1}{N}\sum_{i=1}^N d_{min}(s_i, s^\prime),
\end{equation}
where $d_{min}(s_i, s^\prime)$ is the minimum distance between a point on the traced loop ($s_i$)
and any point on the projected field line ($s^\prime$). Here we consider a number of points on the
traced loop as a function of loop length. We use 1 point for every $4\arcsec$ of loop length, which
gives about 50 points for the longest loops and 4 points for the shortest loops.  An example
calculation illustrating the distances to the two endpoints and the midpoint of the traced loop is
shown in Figure~\ref{fig:mmd}. A very similar metric was used by \citet{savcheva2009} for comparing
flux rope models with soft X-ray images of sigmoids.

By computing the mean minimum distance for each of the candidate field lines we are able to
identify the best-fit or closest field line from the extrapolation for each loop segment. The
field, of course, is continuous and we can only sample it discretely. To increase the probability
that we find the best possible match we have implemented the following procedure for sampling the
domain.  We randomly sample 2000 points on the curtain to use as seeds for computing initial field
lines.  We then choose the best 10 field lines that provide the closest match and consider points
that are slightly perturbed away from the seeds of this first batch. From this we generate a second
batch of candidate field lines and then select the best-fit from all candidates. In total we
consider 4,000 field lines per traced loop.  We have tested this procedure by manually tracing out
projected field lines and supplying them to the algorithm as if they were traced loops. The mean
minimum distance metric for these test loops is typically less than 1\arcsec and the best-fit field
line is always close to the input field line.

One important issue is that the loop tracing does not necessarily return complete loops. Most
traced loops are likely to be only a loop segment sampled from a longer loop. Even if the algorithm
does manage to trace out a complete loop, we wouldn't necessarily be certain of this and be able to
add further constraints on the location of the field line footpoints. Thus it is easy to imagine
scenarios where the closest matched field line is not really related to the traced loop.  A long,
overlying field line, for example, could be matched to a small loop segment that actually lies
close to the solar surface. Unfortunately, it is not obvious how to resolve this limitation. Some
ideas will be discussed in the final section of the paper.

\subsection{Misalignment Angle} \label{sec:angle}

\begin{figure*}[t!]
  \centerline{%
    \includegraphics[width=0.33\textwidth]{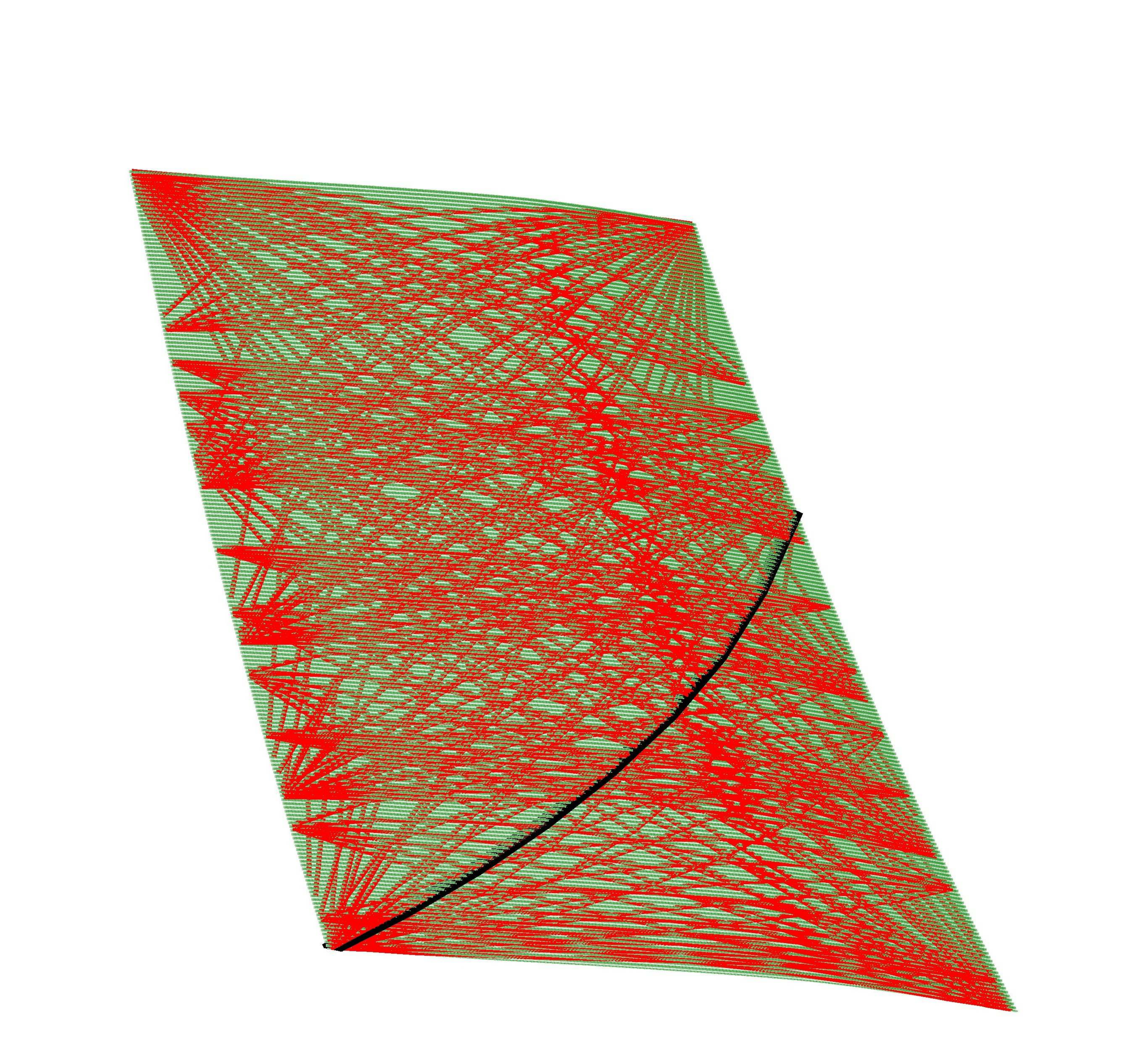}
    \includegraphics[width=0.33\textwidth]{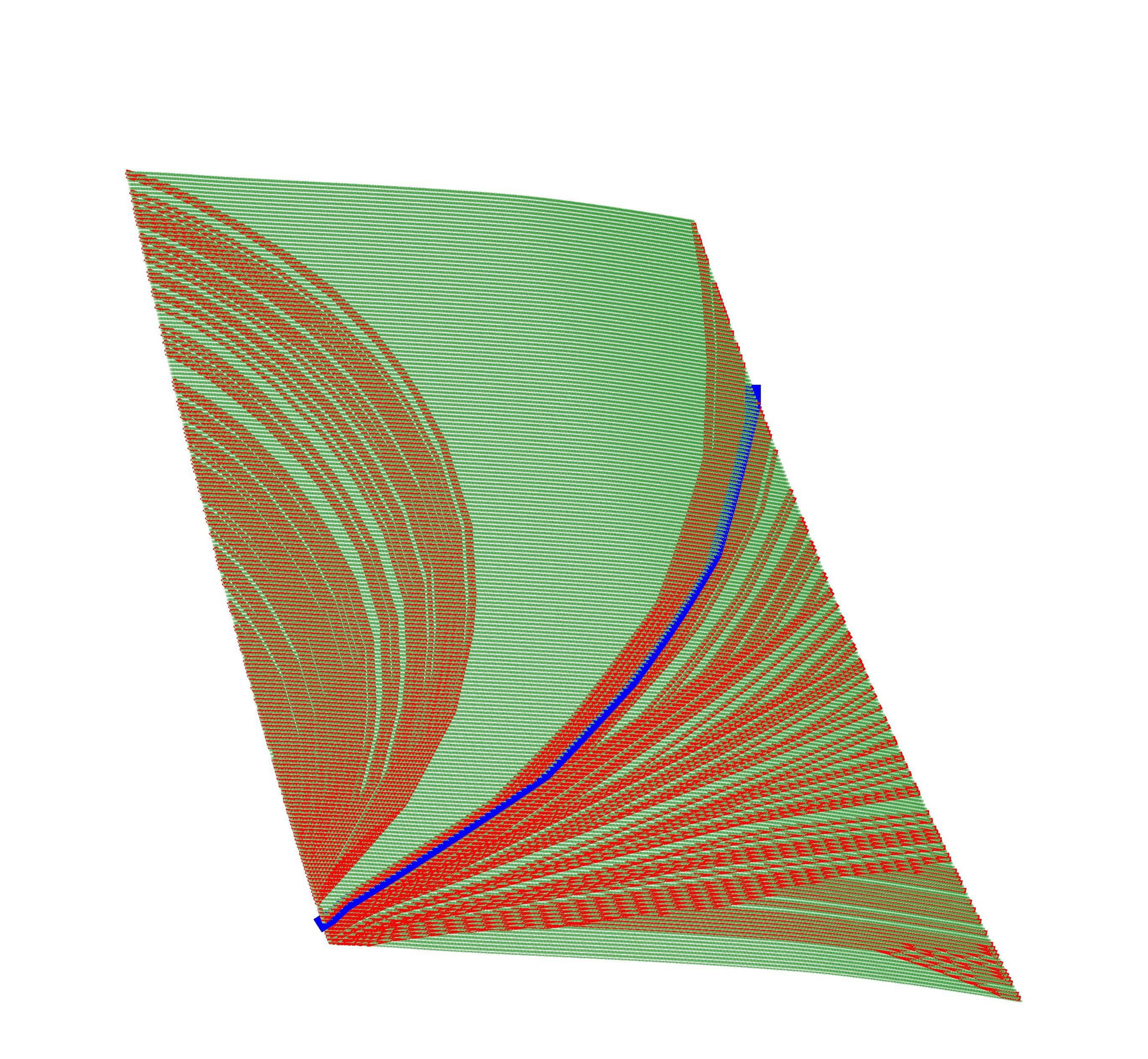}
    \includegraphics[width=0.33\textwidth]{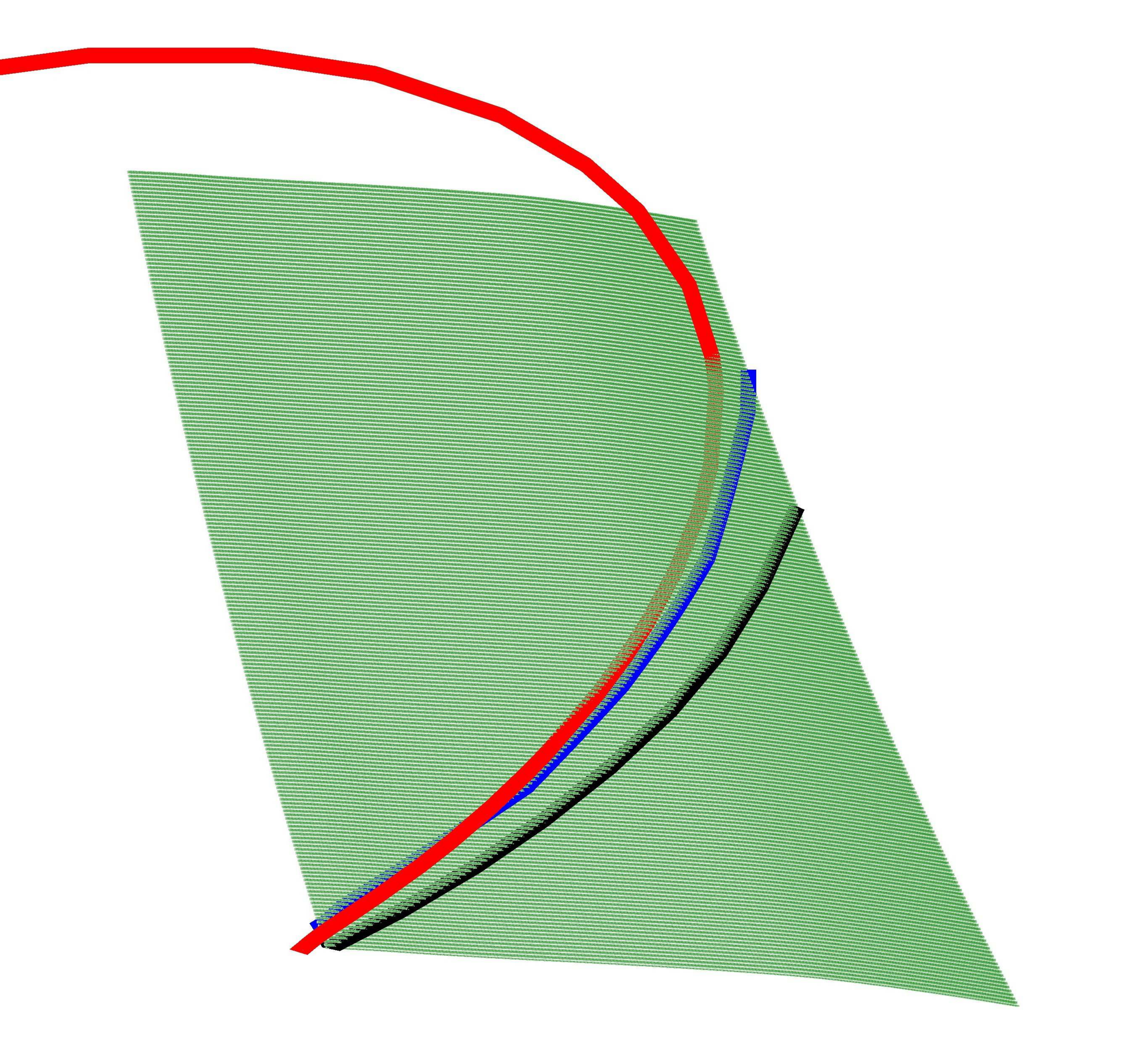}
  }
  \caption{An example of a misalignment angle calculation. Here test geometries are traced across
    the surface of points that project back to the traced loop (this is the example from
    Figure~\ref{fig:curtain1}) and the geometry with the smallest misalignment angle is chosen to
    represent the loop in 3D. Left panel: The 180 test geometries assumed in the VCA model. The
    geometry with the smallest median misalignment angle is indicated in black.  Center panel: Test
    geometries generated by following paths of minimum misalignment angle across the surface. The
    geometry with the smallest median misalignment angle is indicated in blue. Right panel:
    Comparison of the geometries derived from the misalignment angles with the best-fit field line
    derived from the distance metric, which is shown in red.}
  \label{fig:angle}
\end{figure*}

Another point of comparison between traced loops and the magnetic field is the misalignment angle,
that is, the angle between the vector formed by two points on the traced loop segment and the local
magnetic field vector. For a field line the angle between $\mathbf{B}$ and $\mathbf{ds}$ is zero by
construction.  If the misalignment angle is large, magnetic field lines will quickly diverge away
from the traced loop and the field is a poor representation of the loop geometry.

For stereoscopically observed loops the three dimensional position of the loop is known and this
quantity is a very useful measure of how well the field represents the loop. \citet{derosa2009},
for example, computed the misalignment angle for several active region loops imaged from
multiple vantage points and found that none of the NLFF extrapolations could improve on the mean
misalignment angle of the potential model ($\sim24^\circ$).

The VCA model is optimized using the misalignment angle, so this metric is also important to
compute for our study, even though the loops are observed only as two-dimensional projections. For
this case the three dimensional geometry of the loop must be estimated. This is done by assuming a
wide variety of parameterized functions for $h(s)$, the variation of height with distance along the
traced loop segment (see Figure 11 in \citealt{aschwanden2016}). The misalignment angle is computed
for each of these parameterizations and the one with the smallest median angle is selected to
represent the traced loop. An example of this calculation is shown in Figure~\ref{fig:angle}.

The VCA model assumes that the 3D geometry of the loop is circular. It is possible to remove this
restriction by selecting a point on the edge of the curtain and, as is illustrated in
Figure~\ref{fig:angle}, following the path of minimum misalignment angle across it. The curve with
the smallest median angle is selected to represent the traced loop. Curves which do not cross the
curtain are excluded. For a small number of cases no curves that cross the curtain are found. Test
calculations suggest that this method produces 3D loop geometries that more closely match the
best-fit field lines than the circular assumption does, and we compute misalignment angles using
both methods.

We note that the misalignment angle calculation and the mean minimum distance calculations are
closely related. The mean minimum distance calculation finds the field line that is the closest
match to the loop observed on the image plane. The misalignment angle calculation finds the 3D loop
geometry that matches the traced loop in 2D and is most like a field line.
\section{Results} \label{sec:res}

\begin{deluxetable*}{rrrrrrr}
  \tabletypesize{\small}
  \tablecaption{Active Region Summary\tablenotemark{a}}
  \tablehead{
    \multicolumn{1}{c}{Region}      &
    \multicolumn{1}{c}{NOAA}        &
    \multicolumn{1}{c}{Date}        &
    \multicolumn{1}{c}{$X_{cen}$}   &
    \multicolumn{1}{c}{$Y_{cen}$}   &
    \multicolumn{1}{c}{$X_{fov}$}   &
    \multicolumn{1}{c}{$Y_{fov}$}
  }
  \startdata
   1  &  11082  &  19-Jun-2010 01:27:42  &  -308.8  &   470.8  &  322  &  322  \\
   2  &  11082  &  21-Jun-2010 01:16:27  &   112.9  &   423.1  &  392  &  392  \\
   3  &  11089  &  23-Jul-2010 14:32:56  &  -380.5  &  -437.9  &  402  &  402  \\
   4  &  11109  &  29-Sep-2010 23:21:34  &   340.9  &   245.9  &  462  &  462  \\
   5  &  11147  &  21-Jan-2011 13:40:56  &   -61.3  &   470.7  &  502  &  502  \\
   6  &  11150  &  31-Jan-2011 10:55:11  &  -587.7  &  -258.0  &  392  &  392  \\
   7  &  11158  &  12-Feb-2011 15:01:57  &  -306.2  &  -206.8  &  352  &  352  \\
   8  &  11187  &  11-Apr-2011 11:30:35  &  -530.3  &   283.3  &  512  &  512  \\
   9  &  11190  &  15-Apr-2011 00:47:05  &   190.5  &   307.5  &  492  &  492  \\
  10  &  11193  &  19-Apr-2011 13:02:06  &   -13.7  &   372.0  &  492  &  492  \\
  11  &  11243  &  02-Jul-2011 03:08:12  &  -357.2  &   167.9  &  372  &  372  \\
  12  &  11259  &  25-Jul-2011 09:05:57  &   180.9  &   324.3  &  322  &  322  \\
  13  &  11271  &  21-Aug-2011 11:56:09  &   -48.7  &   133.1  &  552  &  552  \\
  14  &  11339  &  08-Nov-2011 18:44:44  &    51.0  &   246.0  &  552  &  552  \\
  15  &  11339  &  10-Nov-2011 11:03:29  &   374.9  &   256.1  &  482  &  482  \\
  \enddata
  \tablenotetext{a}{The times listed are for the HMI line-of-sight magnetograms used in the potential
    extrapolations. The corresponding line-of-sight magnetograms in the Aschwanden extrapolations and
    vector magnetograms in the Wiegelmann extrapolations are within 30 sec and 7 min, respectively,
    of those listed. $X_{cen}$, $Y_{cen}$ are the NOAA active region coordinates of the patch centers
    and $X_{fov}$, $Y_{fov}$ are the width and height of the fields of view in arc-seconds.}
  \label{table:table1}
\end{deluxetable*}

We have described all of the elements that are needed to carry out this study. We have three
different methods for computing the magnetic field components, we can compute field lines and map
them back and forth between the computational domain and the image plane, we can automatically
trace out loops in coronal images, we have a method for matching field lines to traced loops as
well as methods for estimating the misalignment angle. We now turn to the application of this
methodology to an ensemble of active regions.

For this study we use the 15 active regions from \citet{warren2012}, who used these regions to
study the dependence of active region temperature structure on the properties of the magnetic
field. These regions cover about an order of magnitude in the total unsigned magnetic flux,
$4\times10^{21}$ -- $3\times10^{22}$\,Mx., covering almost the full range of typically observed
active regions. Information on these regions is listed in Table~\ref{table:table1}. Note that the
times correspond to the midpoints of raster observations with the EUV Imaging Spectrometer on
\textit{Hinode} (EIS, \citealt{culhane2007}).  For each region we have manually selected a field of
view that includes all of the flux from the active region core (see Figure~\ref{fig:fov}). Using the
field of view and time we downloaded cutouts from the SDO Joint Science Operations
Center\footnote{http://jsoc.stanford.edu/} for a one hour interval beginning with the time listed
in the table. The downloads included all of the AIA EUV channels (171, 193, 211, 335, 94, 131, 304)
at 12\,s, cadence, the AIA UV channels (1600, 1700) at 24\,s cadence, the HMI line of sight
magnetograms at 45\,s cadence, and the HMI vector data at 720\,s cadence. As noted earlier, the
VCA-NLFF code independently downloads single, full-disk AIA EUV and HMI line-of-sight images.

\begin{deluxetable*}{rrrrrcrrrrcrrrr}
  \tabletypesize{\small}
  \tablecaption{Summary of Loop Comparisons}
  \tablehead{
    \multicolumn{1}{c}{} &
    \multicolumn{4}{c}{Minimum Distance\tablenotemark{a}} &
    \multicolumn{1}{c}{} &
    \multicolumn{4}{c}{Misalignment Angle (circular)\tablenotemark{b}} &
    \multicolumn{1}{c}{} &
    \multicolumn{4}{c}{Misalignment Angle (arbitrary)\tablenotemark{c}} \\
    \multicolumn{1}{c}{Wavelength}  &
    \multicolumn{1}{c}{N Loops}  &
    \multicolumn{1}{c}{PFE}      &
    \multicolumn{1}{c}{VCA}      &
    \multicolumn{1}{c}{NLFF}     &
    \multicolumn{1}{c}{}     &
    \multicolumn{1}{c}{N Loops}  &
    \multicolumn{1}{c}{PFE}      &
    \multicolumn{1}{c}{VCA}      &
    \multicolumn{1}{c}{NLFF}     &
    \multicolumn{1}{c}{}     &
    \multicolumn{1}{c}{N Loops}  &
    \multicolumn{1}{c}{PFE}      &
    \multicolumn{1}{c}{VCA}      &
    \multicolumn{1}{c}{NLFF}
  }
  \startdata
         All                  &        12202 &         1.45 &         1.72 &         0.70 &  &          12202 &         11.7 &         11.4 &         12.3 & &          11779 &          5.2 &          5.4 &          4.8  \\
          94                  &          582 &         1.11 &         1.25 &         0.55 &  &            582 &         11.5 &         10.6 &         11.9 & &            568 &          5.2 &          4.6 &          4.5  \\
         131                  &         1357 &         1.20 &         1.52 &         0.63 &  &           1357 &         10.5 &          9.9 &         11.3 & &           1318 &          4.6 &          4.8 &          4.5  \\
         171                  &         2759 &         1.64 &         1.92 &         0.80 &  &           2759 &         11.4 &         10.8 &         12.1 & &           2634 &          5.1 &          5.5 &          4.8  \\
         193                  &         3224 &         1.57 &         1.89 &         0.73 &  &           3224 &         12.1 &         12.2 &         12.9 & &           3115 &          5.5 &          5.8 &          4.9  \\
         211                  &         2997 &         1.50 &         1.79 &         0.70 &  &           2997 &         12.2 &         12.1 &         12.8 & &           2889 &          5.4 &          5.6 &          4.9  \\
         335                  &         1283 &         1.15 &         1.29 &         0.58 &  &           1283 &         11.8 &         10.9 &         11.9 & &           1255 &          5.1 &          4.9 &          4.5  \\
         Hot\tablenotemark{d} &          214 &         1.46 &         1.31 &         0.61 &  &            214 &         13.2 &         11.7 &         13.0 & &            214 &          6.2 &          6.1 &          4.9  \\
    \enddata
  \tablenotetext{a}{The median of the mean minimimum distance in arc-seconds is listed for each
    extrapolation.}
  \tablenotetext{b}{The median of the misalignment angle in degrees is listed for each
    extrapolation. The 3D loop geometry assumed in the VCA is used for all three extrapolations.}
  \tablenotetext{c}{The median of the misalignment angle in degrees is listed for each
    extrapolation. Here arbitrary curves that follow the minimum misalignment angle are used to estimate the
    3D loop geometry. In a small number of cases, no curve was found to cross the curtain and those have
    been excluded from the summary.}
    \tablenotetext{d}{The hot loops category are loops identified with \ion{Fe}{18} emission in the core of the active region.}
  \label{table:table2}
\end{deluxetable*}

For each region we computed the potential, NLFF, and VCA-NLFF magnetic field extrapolations and
saved the field components to a file. Example field line calculations for each region are shown in
Figures~\ref{fig:compare1}--\ref{fig:compare3}. Note that in these plots randomly selected field
lines are shown, they have not been matched to any traced loops. Also, field lines for the same
randomly selected seed points are shown in each row.

\begin{figure*}[t!]
\centerline{\includegraphics[clip, width=0.95\textwidth]{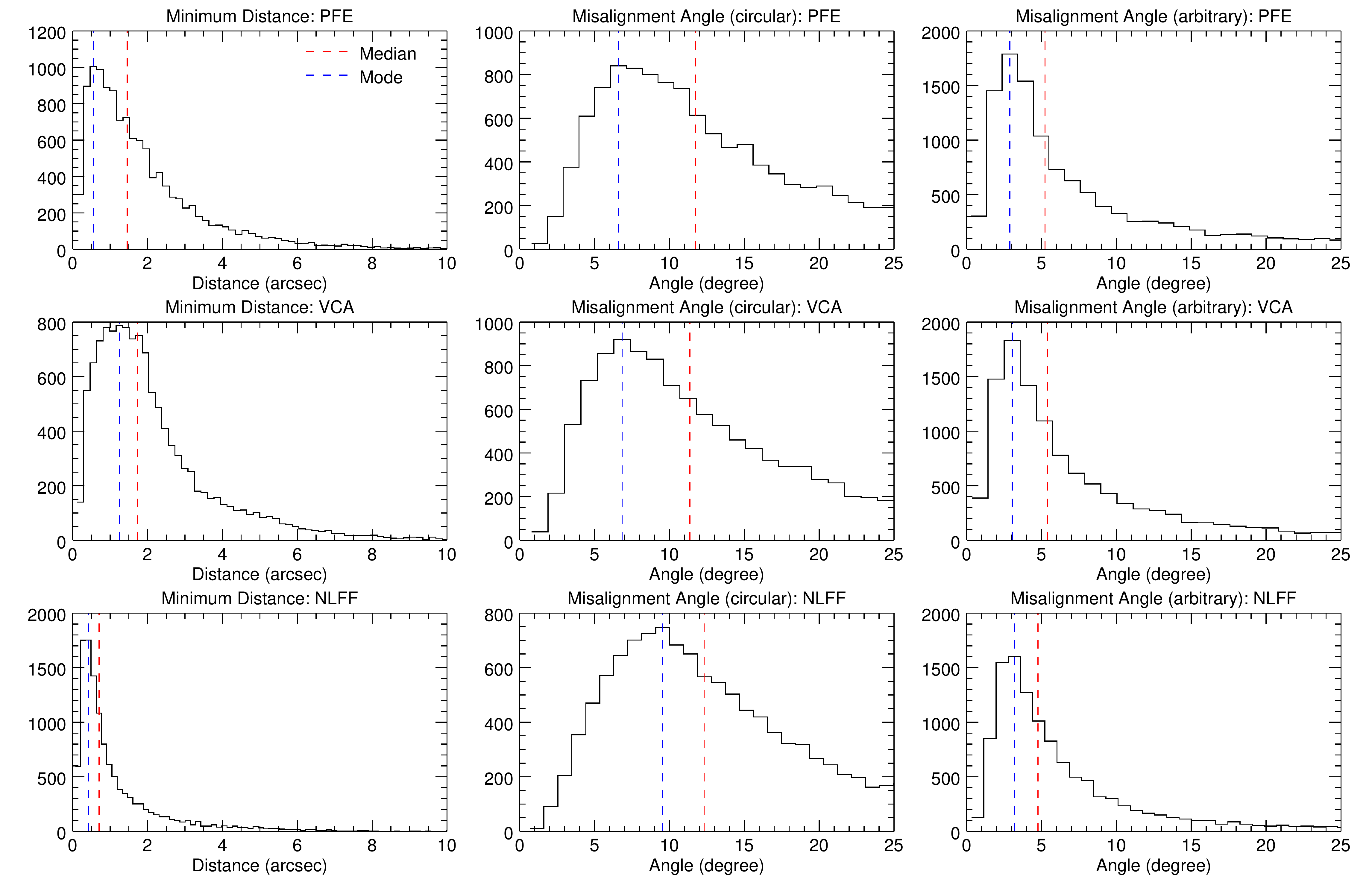}}
\caption{Distributions of minimum distances and misalignment angles for the three extrapolation
  methods. In general, the differences among the distributions are small. The distance metric for
  the vector NLFF shows the closest matches and has the narrowest distribution. }
\label{fig:hist}
\end{figure*}

For each region we traced loops in the AIA 94, 131, 171, 193, 211, and 335 images. A total of
12,202 loop segments were identified. The loop segments range in projected length from
17\arcsec\ to about 200\arcsec. The distribution of loop lengths is a power law with an index of
approximately 3, consistent with \citet{aschwanden2013c}. As is evident in
Figure~\ref{fig:tracing}, the majority of the loop segments are identified in the 171, 193, and 211
channels. These channels generally show emission from ions formed at about 1\,MK and thus these
comparisons are heavily weighted towards loops at this temperature.

{As mentioned in Section~\ref{sec:auto}, we have identified loop segments independently of the
  VCA-NLFF algorithm, but there is some overlap. Of the 12,202 loop segments from our sample, only
  2,613 (about 21\%) were also used in optimizing the VCA-NLFF model parameters, and thus a large
  fraction of the loops used to evaluate its performance are independent of the training data.}

We computed the mean minimum distance metric for each traced loop for each of the
extrapolations. The result of this calculation is summarized in Table~\ref{table:table2}, where we
present median values for all of the loops and for the loops in each AIA wavelength
individually. For this metric, the NLFF extrapolation indicates better fits than both the potential
and VCA models. This is true for both the aggregate value and for each AIA wavelength considered
individually.

As discussed previously, the VCA method is optimized using the misalignment angle and we also
computed this metric for each traced loop for each of the three extrapolations. As indicated in
Table~\ref{table:table2}, the VCA yields smaller misalignment angles than either the potential or
NLFF methods, although the differences are generally small. This calculation assumes that the 3D
loop segments are circular. If we relax the assumption of circular loop segments and consider 3D
loop segments that follow paths that minimize the misalignment angle (see Figure~\ref{fig:angle}),
the median misalignment angle is reduced and the NLFF extrapolation yields the smallest values.

\begin{figure*}[t!]
\centerline{\includegraphics[width=0.95\textwidth]{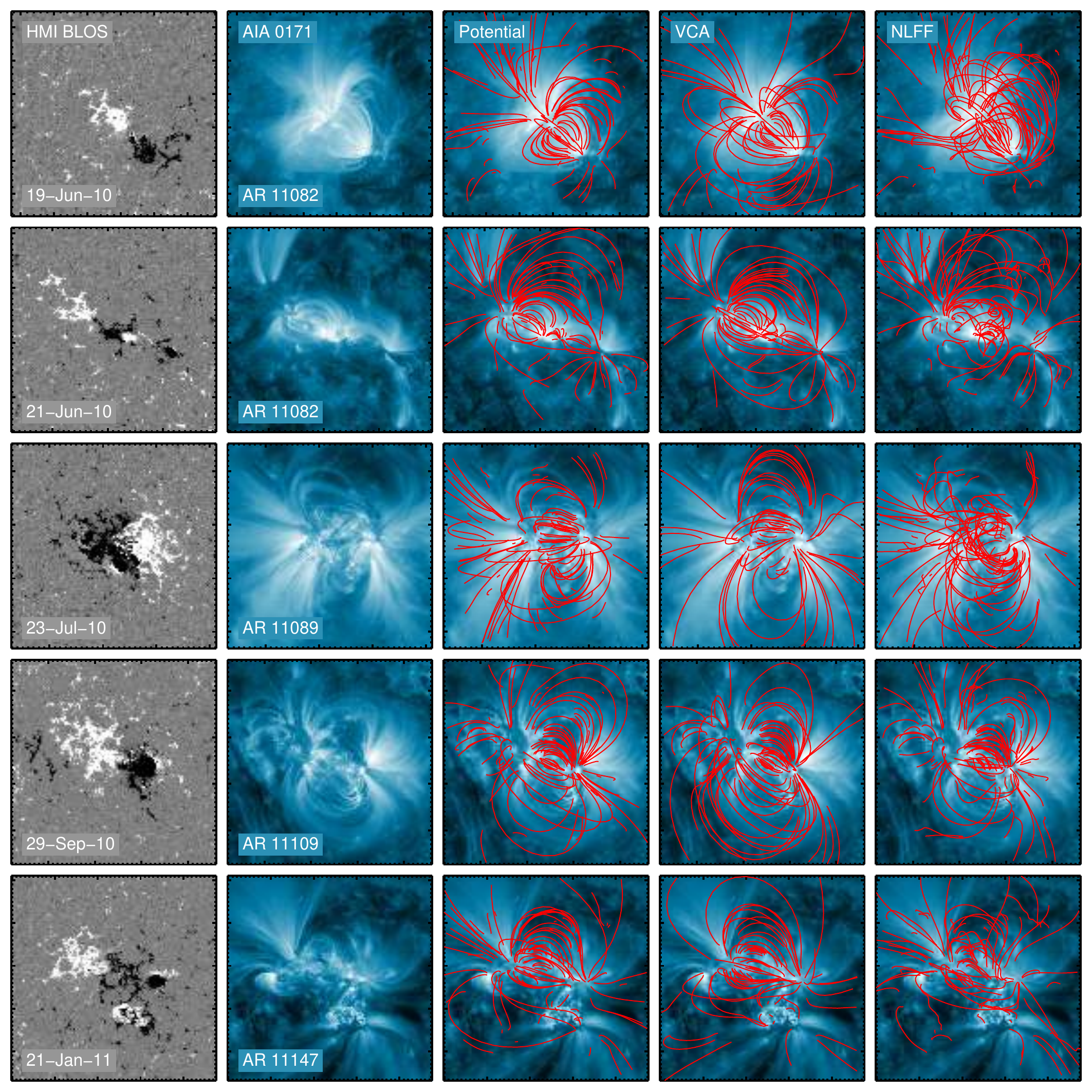}}
\caption{Comparison of extrapolation results. HMI and AIA observations are shown for the active
  regions in this study along with the resulting field lines from each extrapolation.  Field lines
  are computed using common seed points and projected onto the AIA 171 \AA\ image.  {Note that
    these field lines are selected randomly and do not correspond to traced loops.} Active regions
  1--5 are shown.}
\label{fig:compare1}
\end{figure*}

\begin{figure*}[t!]
\centerline{\includegraphics[width=0.95\textwidth]{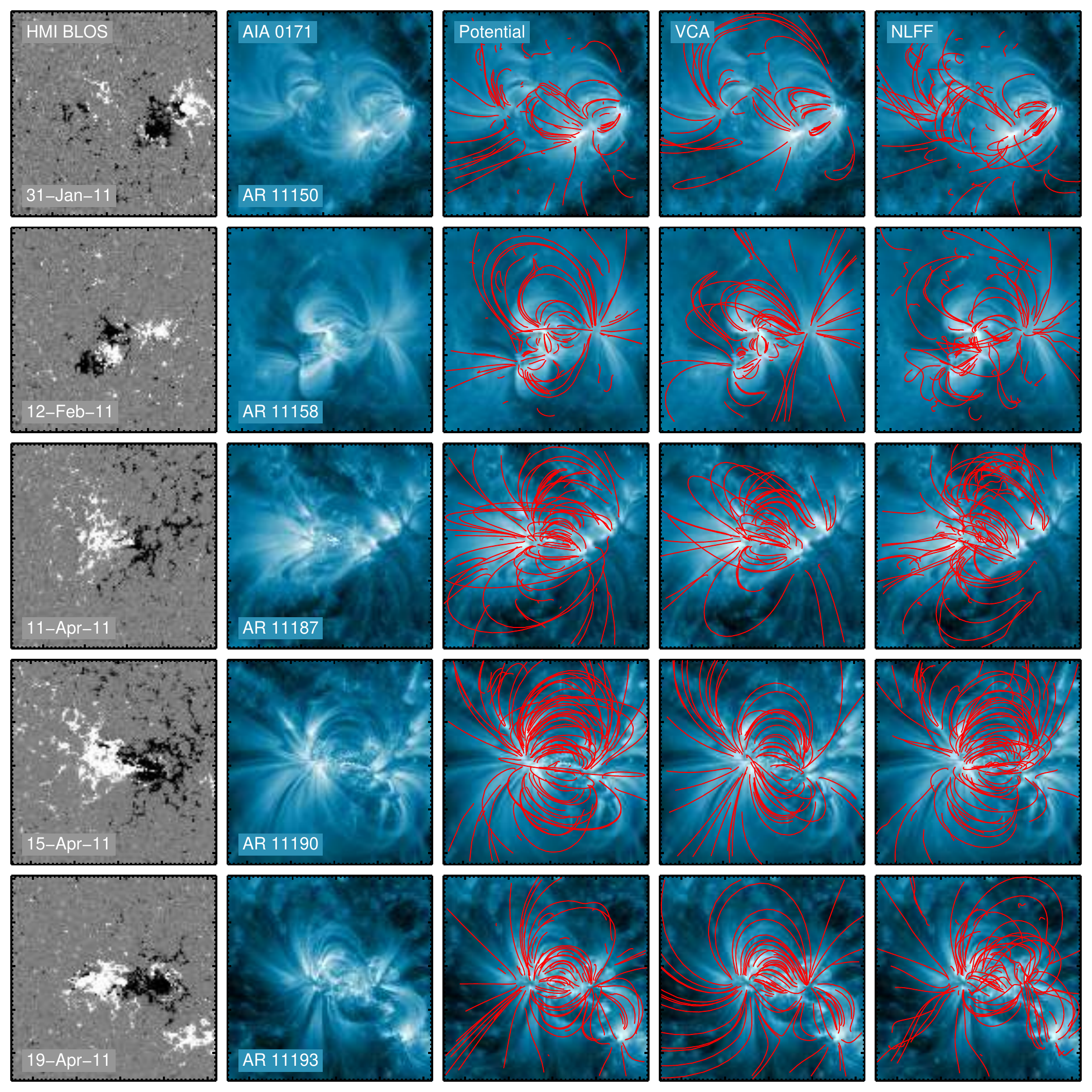}}
\caption{Same as Figure~\ref{fig:compare1} for active regions 6--10.}
\label{fig:compare2}
\end{figure*}

\begin{figure*}[t!]
\centerline{\includegraphics[width=0.95\textwidth]{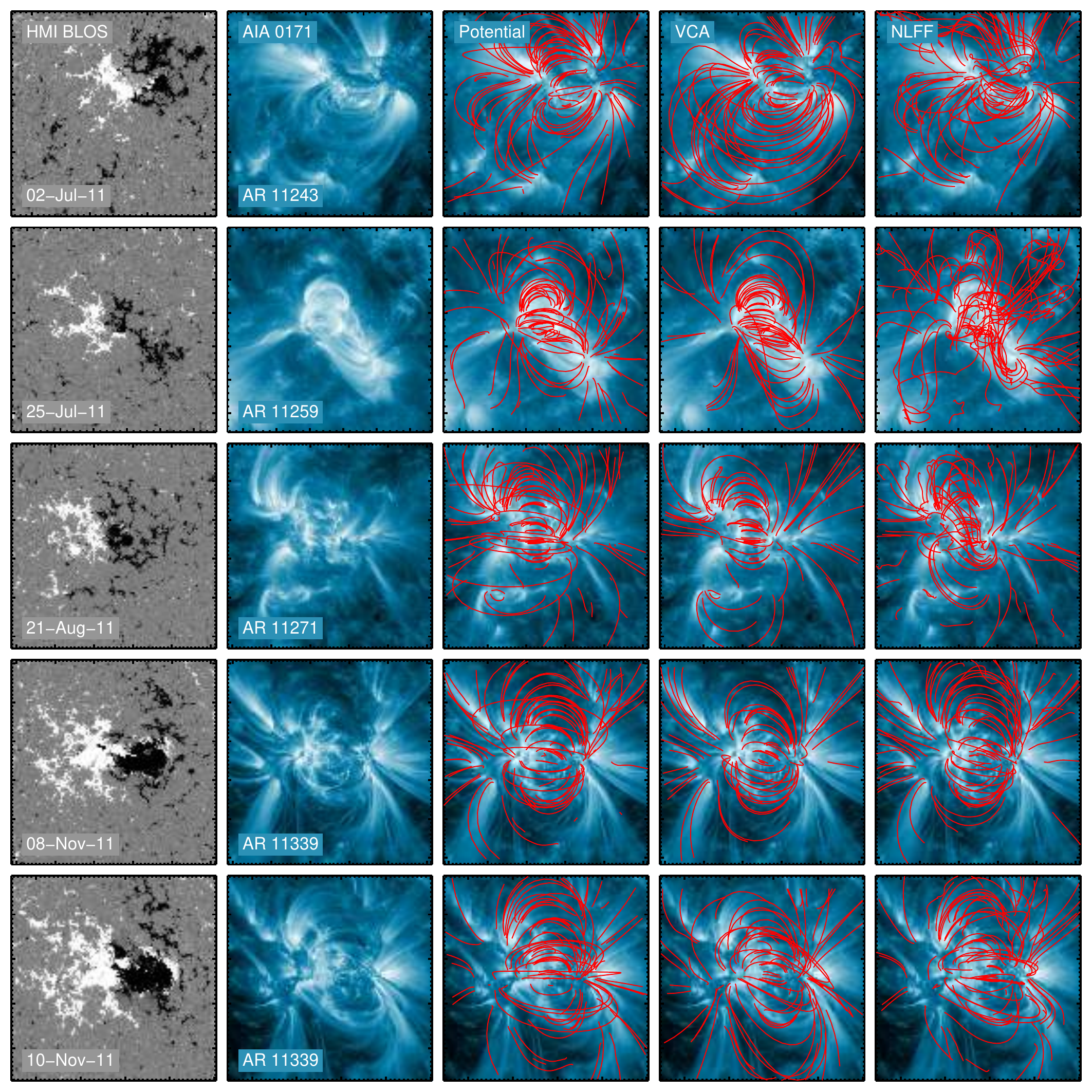}}
\caption{Same as Figure~\ref{fig:compare2} for active regions 11--15.}
\label{fig:compare3}
\end{figure*}

We have also examined the distributions of distances and misalignment angles for each of the
extrapolations. As is shown in Figure~\ref{fig:hist}, these distributions are not Gaussian, but
resemble log-normal or power-law distributions. Thus the median and the mode of each distribution
are not the same. The general trends, however, are consistent with the results summarized in
Table~\ref{table:table2}. The smallest deviations and the narrowest distribution is for the
distance metric applied to the vector NLFF extrapolation. The other metrics and extrapolation
techniques generally yield similar results.

The values that we obtain for the misalignment angle are about a factor of two smaller then what
was presented by \citet{derosa2009} for a set of stereoscopically observed loops. Since we obtain a
consistent misalignment angle of $\approx 10^\circ$--$13^\circ$, independent of the active region,
observed wavelength, or magnetic field code (PFE, VCA, NLFF), as well as the consistency of the
misalignment angles found in other studies (see \citealt{aschwanden2016b} Table 3), we conclude
that the smaller misalignment angle that we find in this study is not due to a data selection
effect, but rather a limitation of the stereoscopic triangulation method using STEREO data.  STEREO
has a much poorer spatial resolution (pixel size or 1.59\arcsec\ and spatial resolution of
$\approx4.0$\arcsec) than AIA (0.6\arcsec\ pixel size and spatial resolution of
$\approx1.5$\arcsec). The stereoscopic error itself was determined to be of order
$7.6^\circ$--$11.5^\circ$ (Table 2 in \citealt{aschwanden2010b}). This leads to misalignment angles
of $19^\circ\pm3^\circ$ for the 3D-misalignment angle (Table 3 in \citealt{aschwanden2013}), or
$14.3^\circ$--$19.2^\circ$ (Section 3.3 in \citealt{aschwanden2012c}). Thus, magnetic field
modeling with AIA data yields typical misalignment angles of $\approx 10^\circ$, while
stereoscopically triangulated loops using STEREO data produce a misalignment angle that is about a
factor of 2 larger.

To further explore the bias towards potential loops imaged in the AIA 171, 193, and 211 channels we
have attempted to isolate loops associated with high-temperature emission. To do this we have
processed the AIA 94 images to remove the contribution from million-degree plasma and isolate the
emission from the \ion{Fe}{18} 93.92\,\AA\ line, which is formed at about 7\,MK (see
\citealt{warren2012} for details). We re-ran the loop tracing algorithm on these processed images
and identified a new set of hot loops. We then matched these loops to the loops from our original
ensemble that used the unprocessed images. Visual inspection shows that these hot loops are
preferentially found in the active region core.

The results from these hot loops are summarized in the final row of Table~\ref{table:table2}. As
expected, for this population of hot loops the potential extrapolation is outperformed by the NLFF
methods in all metrics. The differences, however, are not particularly large. Unfortunately, we are
able to identify only 214 hot loops in our sample of 12,202 total loops.
\section{Summary and Discussion} \label{sec:summary}

We have presented systematic comparisons between magnetic field lines computed from three different
extrapolation methods and the topology of coronal loops inferred from AIA images. The NLFF methods
generally provide better matches between the field and the observed loops. The NLFF method based on
vector data yields the smallest values for the distance metric. The VCA and NLFF methods yield
smaller values for the misalignment angle than the potential. The differences, however, are
generally small: about 1\arcsec\ for the distance metric and about 1$^\circ$ for the misalignment
angle. A visual inspection of the best-fit field lines, such as those presented in
Figures~\ref{fig:curtain1} and \ref{fig:curtain2}, also suggest relatively small differences
between the different extrapolation methods.

This study highlights some fundamental limitations of the available data and extrapolation
methods. Improvements in these areas should lead to better fits between the extrapolations and the
observed loops.

\paragraph{High Temperature Emission} As noted earlier, the majority of the identified loops are
from emission formed at a about 1\,MK. The currents are likely to be strongest along the neutral
line in the active region core, where loops generally have much higher temperatures. Studies with
\textit{Hinode}/EIS have shown that these loops are generally about 4\,MK
\citep{warren2011,warren2012,delzanna2014}. The AIA 94 channel includes \ion{Fe}{18}, but this is
formed at about 8\,MK and strong \ion{Fe}{18} emission is generally only observed in large active
regions or in transient heating events. When \ion{Fe}{18} emission is observed, loop identification
may be improved if the AIA 94 images are processed to remove the contribution from lower
temperature emission (see \citealt{warren2012,teriaca2012}).

It is likely that observations from the X-ray Telescope on \textit{Hinode} (XRT,
\citealt{golub2007}) could be used for identifying high temperature loops. As mentioned previously,
\citet{savcheva2009} used XRT images to constrain a NLFF model of a sigmoid, but selected the
observed loops manually. However, the broad temperature response of XRT may limit the efficacy of
the automated loop tracing.

High spatial resolution observations of emission lines formed at about 4\,MK, such as \ion{Ca}{14}
193.874\,\AA, would be ideal for identifying the topology of loops in the active region core. Such
an instrument is being considered for a future Japanese space mission (e.g.,
\citealt{teriaca2012b}).

\paragraph{Chromospheric Emission} Matching chromospheric structures observed at high spatial
resolution with \textit{IRIS} or a ground-based observatory is a complementary approach to studying
non-potential fields in the core of the active region. Since such data is not available for all of
the active regions considered here, we have not pursued this idea here.  The limited field of view
for high resolution data are also an obstacle to applying such data to this
problem. \citet{aschwanden2016b} has done exploratory calculations for three active regions and was
able to find good agreement between the VCA model and loops traced at chromospheric temperatures
near a sunspot.

\paragraph{Projection Effects} Coronal images show projections of three dimensional structures onto
a two dimensional plane, which limits our ability to compare loops traced in coronal images and
field lines. As we have seen, to project traced loops back to three dimension space involves many
assumptions about the field line geometry. Projecting field lines onto the image plane does not
involve any assumptions, but since we cannot be sure that we are comparing with a complete loop, it
does not yield a unique result.

Observations from multiple viewing angles are an obvious solution to this problem and the
\textit{STEREO} \citep{kaiser2008} mission has provided several examples of this
\citep[e.g.,][]{aschwanden2012a, aschwanden2012b, aschwanden2013, chifu2015,
  chifu2017}. Unfortunately, stereoscopic observations have been very limited and are not likely to
be taken routinely in the near future. Solar Orbiter \citep{muller2013} will take coronal images
from vantage points away from the Sun-Earth line, but the launch of this mission is still several
years away.

One possibility for reducing projection ambiguities is to consider time sequences of images rather
than individual snapshots. The time sequences would allow transient brightenings to be
detected. Since it is likely that the brightening occurs over the entire loop, this would provide
the full loop geometry. This constrains the search space for potential field line matches
considerably. We are currently investigating this approach using observations from the AIA 94
channel.

\paragraph{Preprocessing} \citet{chifu2017} extended the NLFFF optimization code by implementing
the additional constraint to minimize the angle between the reconstructed local magnetic field
direction and the orientation of 3D-loops.  In that study a number of 3D-loops have been
stereoscopically reconstructed from EUV-images from three vantage points (STEREO A, B and SDO). The
method was dubbed S-NLFFF. While in the current implementation the method requires 3D- loops to
constrain the NLFFF-code, a generalization towards using traced 2D-loops from one image is ongoing
work.

\paragraph{Metrics} We have considered two metrics for comparing field lines to observed
loops, the minimum distance and the misalignment angle. When only 2D projected loop observations
are available, the minimum distance metric is likely to be the most useful. This metric identifies
the topological feature of interest (the field line) using the observations directly and avoids the
intermediate step of estimating the 3D loop geometry. Future studies involving 2D observations
should use this metric along with the misalignment angle.


\acknowledgments HPW, NAC, and IUU were supported by the Chief of Naval Research, NASA's
Heliophysics Grand Challenges Research program, and NASA's \textit{Hinode} program. MJA was
partially supported by NASA’s STEREO program. TW acknowledges DFG-grant WI 3211/5-1. We thank the
referee for the careful review and thoughtful comments that helped to improve the paper.


\end{document}